\pdfoutput=1
\documentclass[twocolumn,preprintnumbers,superscriptaddress,amsmath,amssymb,prb]{revtex4-1}

\usepackage{graphicx}
\usepackage{lineno}
\usepackage{dcolumn}
\usepackage{bm}
\usepackage[utf8]{inputenc}
\usepackage{comment}
\usepackage[unicode=true, bookmarks=false, breaklinks=false, pdfborder={0 0 0},colorlinks=false]{hyperref}
\usepackage{breakurl}
\usepackage{url}
\usepackage{natbib}
\usepackage{multirow}
\usepackage{float}
\usepackage{wrapfig}
\usepackage{xcolor}

\begin{document}

\title{A Real-Space Framework for the Bare and Interacting Longitudinal
       Susceptibility of Solids}

\newcommand{\Uppsala}{Department of Physics and Astronomy, Uppsala University, Box 516, SE-75120 Uppsala, Sweden}
\newcommand{\Netherlands}{Institute for Molecules and Materials, Radboud University, NL-6525 AJ Nijmegen, The Netherlands}
\newcommand{\KTH}{Department of Applied Physics, School of Engineering Sciences, KTH Royal Institute of Technology, AlbaNova University Center, SE-10691 Stockholm, Sweden}
\newcommand{\WISEUppsala}{Wallenberg Initiative Materials Science for Sustainability, Uppsala University, 75121 Uppsala, Sweden}
\newcommand{\WISEKTH}{Wallenberg Initiative Materials Science for Sustainability (WISE), KTH Royal Institute of Technology, SE-10044 Stockholm, Sweden}
\newcommand{\SeRC}{SeRC (Swedish e-Science Research Center), KTH Royal Institute of Technology, SE-10044 Stockholm, Sweden}

\author{Rohit Pathak}
    \email[Corresponding author: ]{rohit.pathak@physics.uu.se}
    \affiliation{\Uppsala}
    \affiliation{\WISEUppsala}

\author{Abhay Mishra}
    \affiliation{\Uppsala}

\author{Mikhail I. Katsnelson}
    \affiliation{\WISEUppsala}
    \affiliation{\Netherlands}

\author{Anna Delin}
    \affiliation{\KTH}
    \affiliation{\WISEKTH}
    \affiliation{\SeRC}

\author{Olle Eriksson}
    \affiliation{\Uppsala}
    \affiliation{\WISEUppsala}

\author{Vladislav Borisov}
    \affiliation{\Uppsala}
    \affiliation{\WISEUppsala}

\date{\today}

\begin{abstract}
  We present an implementation of the bare and interacting longitudinal
  susceptibility in a full-potential linear muffin-tin orbital code.
  The bare susceptibility is built in real space from Matsubara sums of
  Kohn--Sham Green's functions and is resolved by site, spin and orbital degrees of freedom; the
  spin and charge susceptibilities follow from a Dyson equation, with an on-site exchange-correlation kernel.
  After testing the solver on the Stoner-like instability of bcc Fe, we apply
  it to four density-wave materials. The bare susceptibility peaks at $H$ in
  bcc Cr, where the nonmagnetic state is close to a Stoner instability, and
  at the stripe vector $M$ in FeSe. In LaTe$_3$ the response of the Te nets peaks close to the observed charge-density-wave vector, at its image in the neighbouring Brillouin zone; which of the two equivalent vectors carries the response is set by the phases of the sites in the cell. In 2H-NbSe$_2$ the response is weakly modulated, and the observed vector lies on the flank of a broad maximum, consistent with the known importance of electron-phonon coupling there. Antiferromagnetic order reduces the bare longitudinal response at the ordering vector by $\sim8\%$ in Cr and about half in FeSe. The orbital decomposition of
  the bare susceptibility shows which orbitals contribute the most for the nearest-neighbor bonds.
\end{abstract}

\maketitle
{\noindent\small\raggedright $^{\ast}$Corresponding author: Rohit Pathak\\ Email: rohit.pathak@physics.uu.se\par}
\medskip

\section{Introduction}
\label{sec:intro}

One of the most striking manifestations of collective electron behaviour
in metals and other itinerant-electron systems is the spontaneous
breaking of translational symmetry.
Rather than remaining spatially homogeneous, the electrons can organise
collectively into periodically modulated ground states known as charge
density waves (CDWs) or spin density waves (SDWs), where the local
charge or spin density develops a regular spatial pattern at a
characteristic wave vector $\mathbf{q}^*$, which may or may not be commensurate
with the underlying lattice~\cite{Gruner1988,Overhauser1962,Zhu2015}.
Understanding the microscopic origin of these modulations is important
both for fundamental physics and for the practical goal of predicting
and controlling material properties.

Density-wave phases compete and cooperate with other electronic orders,
most notably superconductivity in the cuprates
\cite{Keimer2015}, the iron-based superconductors \cite{Fernandes2022},
and the transition-metal dichalcogenides \cite{Rossnagel2011}, and can coexist or intertwine with nematic order
\cite{Fernandes2014}.
The electronic response can also have pronounced orbital character \cite{deMedici2014,Yi2019}, with particular orbital sectors contributing much more strongly than others. Retaining this orbital information is therefore important in a microscopic description of density-wave physics\cite{Johnston2010,Bohmer2018}.

For both CDW and SDW instabilities, a key quantity is the electronic susceptibility $\chi(\mathbf{q})$, which may exhibit a strong enhancement (or in the ideal one-dimensional Peierls limit, a divergence) at the wave vector $\mathbf{q}^*$ connecting nested regions of the Fermi surface, the so-called Fermi surface nesting condition \cite{Peierls1955,Overhauser1962}.
In the ideal one-dimensional Peierls limit, for perfect nesting, the bare (noninteracting) susceptibility
$\chi^0(\mathbf{q})$ develops a logarithmic singularity, and even an
arbitrarily small electron-electron or electron-phonon interaction
suffices to trigger a density-wave instability at the nesting vector.
In real materials, the Fermi surfaces are rarely perfectly nested, so the
interplay between the geometry of the Fermi surface and the strength
and momentum dependence of the relevant interaction determines which, if
any, density-wave order emerges \cite{Rossnagel2011,Johannes2006}.

Fermi surface nesting is, however, only one of several routes to a
density-wave instability. Van Hove singularities near the Fermi level, momentum-dependent electron-phonon coupling, and interband excitonic effects can each
select the ordering vector, sometimes in competition with nesting and
sometimes in its near absence \cite{Irkhin2001,Johannes2006,Weber2011,Zhu2015}.
The static electronic susceptibility computed here contains the effects of the underlying band dispersion, including Fermi surface geometry and van Hove singularities \cite{Irkhin2001}, while retaining the associated orbital and matrix-element information, but it does
not by itself contain the electron-phonon coupling or the electron-hole interaction responsible for an excitonic instability.
One aim of the present work is precisely to separate what the electronic
response alone determines from what it does not: where the bare
$\chi^0(\mathbf{q})$ already peaks at the experimental wave vector, 
the electronic structure itself provides a natural mechanism for selecting the ordering vector; where it does not, additional momentum-dependent interactions or other mechanisms are required to account for its selection.

{Itinerant spin-density-wave order is usually described within a random phase approximation (RPA) in the Bohm--Pines sense \cite{BohmPines1953}, in which the spin density induced at wave vector $\mathbf{q}$ responds to the self-consistent exchange-correlation field at the same $\mathbf{q}$, with contributions from other wave vectors averaged out. The bare susceptibility of the nonmagnetic state is then enhanced by the Stoner factor, and the SDW appears as an instability of this reference state at the wave vector where the enhancement diverges first as the kernel strength grows, rather than being present in the input electronic structure.}
The standard example is elemental bcc chromium, where the near-perfect
nesting between the electron jack-shaped Fermi surface sheet centred at $\Gamma$ and the hole-octahedron Fermi surface sheet centred at $H$ produces an incommensurate SDW along the
$\langle100\rangle$ directions below the N\'eel temperature
$T_N = 311$~K \cite{Overhauser1962,Fawcett1988,Lomer1962}.
Iron-based superconductors, such as BaFe$_2$As$_2$ and FeSe, show a more complex picture.
Their stripe-type magnetic fluctuations persist, and may even strengthen, when
long-range magnetic order is suppressed by chemical doping or applied
pressure~\cite{Bohmer2018,Johnston2010,Wang2016}. These fluctuations are discussed as a
candidate pairing interaction~\cite{Kreisel2020}, which places the
spin susceptibility at the centre of the theoretical description of these materials. One should keep in mind, however, that the RPA description --- being accurate enough for prediction of the ground state ---  can strongly overestimate $T_N$ since within RPA the only way to lose the magnetic order is to suppress the magnetic moment itself, which costs much more energy than disordering the directions of the magnetic moments. The latter proceeds through spin fluctuations that are coupled to each other, which is an effect beyond the RPA. 
At finite temperature, this mode coupling renormalises the Stoner enhancement and sets the ordering temperature on the scale of the inter-site exchange rather than of the exchange splitting~\cite{moriya_book}.

Charge-density waves, while similarly rooted in Fermi surface
instabilities, are further shaped by momentum-dependent electron-phonon
coupling and, in some materials, by interband excitonic condensation.
Transition-metal dichalcogenides (TMDs) such as 1T-TiSe$_2$ and
2H-NbSe$_2$ have served as central testing grounds for the competing
CDW mechanisms for more than four decades
\cite{Wilson1975,DiSalvo1976,Rossnagel2011}.
More recently, the rare-earth tritelluride family, of which LaTe$_3$ is
a representative member, has emerged as a nearly ideal
quasi-two-dimensional platform for exploring CDW formation associated with
imperfect Fermi surface nesting within the nearly square Te-layer planes
\cite{Gweon1998,Brouet2008,Ru2008}, although first-principles work assigns
the ordering vector to momentum-dependent electron-phonon coupling rather
than to nesting itself \cite{Hong2022}.

Many-body perturbation theory, using a Feynman diagram approach, has previously used particle-hole bubbles built from Green's functions in real space. 
For instance, 
Samolyuk and Antropov computed $\chi_{ij}$ from
atomic-sphere Green's functions, with the sum rule
$\sum_j\chi_{ij}=N_i(\varepsilon_F)$, to analyse the magnetic instabilities of
CaFe$_2$As$_2$~\cite{Samolyuk2009}. To our knowledge, a full-potential implementation that keeps the orbital
structure of the real-space particle-hole bubble and treats the longitudinal spin and charge
channels in the same Dyson equation has not been reported.
Most existing approaches either rely on simplified tight-binding models,
sacrifice orbital resolution by evaluating the susceptibility purely in
reciprocal space, or employ constant-matrix-element approximations that
are known to introduce significant errors for transition-metal $d$
electrons \cite{Heil2014,JohannesMazin2008}.
In particular, the widely used ``nesting-function'' evaluation of the bare
susceptibility with constant matrix elements can produce structure that the
full response does not have, since the orbital and matrix-element structure
often reshapes $\chi^0(\mathbf{q})$ substantially relative to the pure
Lindhard geometry \cite{JohannesMazin2008,Heil2014}.
Whether an instability is electronically driven at all is therefore not
settled by Fermi surface geometry alone, and a framework that retains the
matrix elements is needed as much to rule nesting out as to confirm it.
Recent open-source toolkits address part of this problem by evaluating
the Lindhard susceptibility from Kohn-Sham eigenstates with
orbital-dependent form factors \cite{Ekuma2025NESTOR}, but they provide
the bare response only in reciprocal space and do not treat the
interacting spin and charge channels on the same footing.

In this work, we address this by developing a real-space formulation of the bare
susceptibility within the full-potential linear muffin-tin
orbital (FP-LMTO) method as implemented in the RSPt package
\cite{Wills2010}.
The noninteracting susceptibility is assembled from site-resolved
Matsubara Green's functions via a sum over Matsubara frequencies,
retaining the full orbital structure of the retained orbital subspace.
The interacting response is then obtained through an RPA-like Dyson
equation following the formalism of Katsnelson and Lichtenstein
\cite{katsnelson2004magnetic}, and Fourier transformation to reciprocal
space yields $\chi(\mathbf{q})$ along any desired high-symmetry path.
We apply the approach to four benchmark systems spanning qualitatively
different physical situations: the CDW materials 2H-NbSe$_2$
\cite{Wilson1975,Johannes2006,Weber2011} and LaTe$_3$
\cite{Brouet2008,Ru2008,Zong2019}, and the spin-density-wave metal bcc~Cr \cite{Overhauser1962,Fawcett1988,Lomer1962},
and FeSe \cite{Bohmer2018,Wang2016,Hsu2008}, in which stripe spin fluctuations
are strong but do not condense into long-range order at ambient pressure.
For bcc~Cr the bare $\chi^0(\mathbf{q})$ has its maximum at $H$, the
commensurate approximant of the observed incommensurate spin-density-wave vector, and
for FeSe at $M$, the wave vector of the stripe spin fluctuations observed by neutron scattering, which do not order at ambient pressure~\cite{Wang2016}.
For the two CDW compounds the picture is more varied. In LaTe$_3$ the response of the Te nets has its maximum close to the observed modulation, near $\tfrac57\,(2\pi/c)$, the image of $\mathbf{Q}_\mathrm{CDW}$ in the neighbouring Brillouin zone and the wave vector that nests the Fermi surface of a single Te net~\cite{Brouet2008,Yumigeta2021}. A Lindhard calculation, in which the matrix elements are held constant and the phases of the sites are ignored, takes the same value at the two vectors and peaks away from them~\cite{Hong2022}. In 2H-NbSe$_2$ the response is weakly modulated, with several maxima of comparable height, and $\mathbf{Q}_\mathrm{CDW}$ lies on the flank of the largest one, in line with the absence of a nesting singularity~\cite{Johannes2006,Weber2011}. In both compounds the CDW is attributed to momentum-dependent electron-phonon coupling~\cite{Weber2011,Hong2022,Singh2024}.
The orbital character is identified from the projected density of states
and from the orbital-resolved bare susceptibility.

The remainder of the paper is organised as follows.
Section~\ref{sec:theory} describes the theoretical formalism and
computational details.
Results for CDW and SDW systems are presented and discussed in
Sections~\ref{sec:cdw} and~\ref{sec:sdw}, respectively, and the
orbital character of the instabilities is analysed in
Section~\ref{sec:orbital}.
We summarise our findings in Section~\ref{sec:summary}.

\section{Theory and Computational Methods}
\label{sec:theory}

\subsection{Bare Susceptibility from Matsubara Green's Functions}
\label{subsec:chi0}

The fundamental quantity in our approach is the real-space,
spin-resolved bare susceptibility for a pair of lattice sites separated
by the lattice vector $\mathbf{R}$.
It is obtained by summing the product of two single-particle Green's
functions over the fermionic Matsubara frequencies, in the real-space Green's-function
formalism of Ref.~\onlinecite{LKAG1987}:
\begin{equation}
  \chi_{\sigma}^{0}(\mathbf{R}) =
  -T \sum_{n} \operatorname{Tr}\!\left[
    \mathbf{G}_{\sigma}^{0\mathbf{R}}(i\omega_{n})\,
    \mathbf{G}_{\sigma}^{\mathbf{R}0}(i\omega_{n})
  \right],
  \label{eq:chi0}
\end{equation}
where $T$ is the electronic temperature, set by the thermal (Fermi)
smearing width $W = k_{B}T$ used for the Brillouin zone (BZ) integration,
$\omega_n = (2n+1)\pi T$ are the fermionic Matsubara frequencies,
and $\mathbf{G}_{\sigma}^{0\mathbf{R}}(i\omega_n)$ is the
orbital-space Green's function matrix connecting site $0$ to
site $\mathbf{R}$ for spin channel $\sigma$.
The trace runs over the orbital indices included in the susceptibility calculation 
(typically the $d$-shell subspace of the transition-metal atom)
within the muffin-tin spheres.
This real-space formulation separates the site-diagonal
($\mathbf{R}=\mathbf{0}$, on-site) and site-off-diagonal
($\mathbf{R}\neq\mathbf{0}$, intersite) contributions to the
susceptibility.
The on-site term is independent of $\mathbf{q}$ and only sets the
overall baseline (Sec.~\ref{subsec:ft}). The intersite terms carry all
of the $\mathbf{q}$-dependence once Fourier transformed, and it is these
that determine ultimately where $\chi(\mathbf{q})$ peaks.

The spin-channel-resolved bare susceptibility $\chi^0_\uparrow(\mathbf{R})$
and $\chi^0_\downarrow(\mathbf{R})$ obtained from Eq.~\eqref{eq:chi0}
corresponds physically to the Lindhard particle-hole bubble, describing the response of noninteracting Kohn--Sham quasiparticles to the creation of single particle-hole excitations, for each
spin channel separately.
Their sum $\chi^0_\uparrow(\mathbf{R}) + \chi^0_\downarrow(\mathbf{R})$
gives the total bare susceptibility $\chi^0(\mathbf{R})$, whose Fourier
transform at $\mathbf{q}=\mathbf{0}$ is close to the density of states at the Fermi level of the retained orbitals [Eq.~\eqref{eq:sumrule}], which provides an approximate check on the calculation.

Retaining the orbital indices of the two Green's functions in
Eq.~\eqref{eq:chi0} instead of tracing over them yields an
orbital-resolved bare susceptibility $\chi^0_{\sigma,LL'}(\mathbf{R})$,
the sum of whose matrix elements over both orbital indices returns the
scalar bubble of Eq.~\eqref{eq:chi0}.
The diagonal elements measure the contribution of individual orbitals,
while the off-diagonal elements quantify interorbital (orbital-mixing)
contributions; we use this decomposition to identify the orbital
character of each instability (Sec.~\ref{sec:orbital}).

This decomposition refers to a particular set of local orbitals and is therefore
basis dependent, unlike the trace in Eq.~\eqref{eq:chi0}.
We use the muffin-tin head projection of RSPt,
where the local orbitals are the normalised radial solutions inside the muffin-tin sphere~\cite{Grechnev2007}. Trace quantities
such as $\chi^0_\sigma(\mathbf{q})$ do not depend on the orientation of the
local axes, but they do depend on which orbitals are projected; the sum rule
of Sec.~\ref{subsec:ft} tests this.

\subsection{Interacting Susceptibility via the RPA Dyson Equation}
\label{subsec:dyson}

The interacting longitudinal spin susceptibility $\chi_{zz}(\mathbf{q})$
and charge susceptibility $\chi_{nn}(\mathbf{q})$\cite{katsnelson2004magnetic} are obtained from a
random phase approximation (RPA) that dresses the bare response with a
local (site-diagonal, momentum-independent) electron-electron
interaction kernel $\mathbf{U}$, following the adiabatic time-dependent
density-functional formalism of Katsnelson and
Lichtenstein~\cite{katsnelson2004magnetic}.
Within spin-density functional theory, in the local spin-density approximation or, as here, the generalised gradient approximation, this kernel is not an
external parameter of a model Hamiltonian but is fixed by the
exchange-correlation functional, its elements being, up to an overall
sign, the second derivatives of the exchange-correlation energy with
respect to the spin-resolved densities
[cf.\ Ref.~\onlinecite{katsnelson2004magnetic}, Eq.~(14)],
\begin{equation}
  U_{\sigma\sigma'} =
  -\,\frac{\partial^2\!\left(n\,\varepsilon_{\mathrm{xc}}\right)}
       {\partial n_\sigma\,\partial n_{\sigma'}},
  \qquad
  n_\sigma = \tfrac{1}{2}\!\left(n + \sigma m\right),
  \label{eq:Ukernel}
\end{equation}
so that $\mathbf{U}$ is a $2\times2$ matrix in spin space. The second
derivatives are negative, because exchange and correlation are
attractive; the minus sign makes $\mathbf{U}$ positive, like the bare
response at $\mathbf{q}=\mathbf{0}$ [Eq.~\eqref{eq:sumrule}]. $\mathbf{U}$
is therefore the exchange-correlation kernel, not a screened Coulomb or
Hubbard interaction, and is not to be compared with Hubbard or
constrained-RPA values of $U$. A local, $\mathbf{q}$-independent kernel is
supported by Sandratskii and K\"ubler, who found the Stoner parameter of
V, Cr, Rh and Pd in the local density approximation to depend only weakly
on $\mathbf{q}$~\cite{Sandratskii1992}. We have evaluated
Eq.~\eqref{eq:Ukernel} numerically for all studied systems as a weighted average of the second derivative of the exchange-correlation energy over each muffin-tin sphere, with the valence spin density as the weight [Eq.~\eqref{eq:kernel_def}], and list the calculated values as well as further technical details in Appendix~\ref{app:U}. 
We also consider $\mathbf{U}$ as an on-site
free parameter and verify how the interacting charge and spin susceptibilities change upon variation of $U$ to understand and visualise the effect of correlations (Sec.~\ref{subsec:compdet}). The tabulated Stoner
integrals of Refs.~\onlinecite{Janak1977,Gunnarsson1976} measure the
exchange splitting per unit moment and are therefore to be compared with
the Stoner parameter $I$ of Eq.~\eqref{eq:stoner}, not with the individual
elements $U_{\sigma\sigma'}$.

To construct the interacting response, the spin-resolved real-space
bubbles are first Fourier transformed and assembled into the diagonal
bare susceptibility matrix
\begin{equation}
\boldsymbol{\chi}^{0}(\mathbf{q}) =
\begin{pmatrix}
\chi^{0}_{\uparrow}(\mathbf{q}) & 0 \\[4pt]
0 & \chi^{0}_{\downarrow}(\mathbf{q})
\end{pmatrix},
\label{eq:chi0mat}
\end{equation}
which is diagonal because the bare response has no cross-spin component in
a collinear state, nonmagnetic or ordered; $\chi^0_\sigma(\mathbf{q})$ is
the Fourier transform over bond vectors given in Eq.~\eqref{eq:ft} below,
and the interacting response is obtained by solving the Dyson equation
\begin{equation}
  \mathbf{K}(\mathbf{q}) =
  \left[\mathbf{1} - \boldsymbol{\chi}^0(\mathbf{q})\,
    \mathbf{U}\right]^{-1} \boldsymbol{\chi}^0(\mathbf{q}),
  \label{eq:dyson}
\end{equation}
where $\mathbf{1}$ is the $2\times2$ identity matrix and
$\mathbf{K}(\mathbf{q})$ is the interacting response matrix, whose element
$K_{\sigma\sigma'}$ gives the change in the spin-$\sigma$ density produced by a
perturbation acting on spin $\sigma'$ [see Ref.~\onlinecite{katsnelson2004magnetic}, Eq.~(15)].
Reference~\onlinecite{katsnelson2004magnetic} writes $\mathbf{X}$ for the bare matrix
that we denote $\boldsymbol{\chi}^0$; we keep a single symbol for the bare
response throughout.
Equation~\eqref{eq:dyson} is the compact matrix form of the coupled response equations of Ref.~\onlinecite{katsnelson2004magnetic} [see Eq.~(19) in this reference], obtained
by Fourier transforming the operator relation
$\mathbf{K} = \boldsymbol{\chi}^0 + \boldsymbol{\chi}^0\mathbf{U}\mathbf{K}$ for a local
kernel $\mathbf{U}$; the four scalar components
$K_{\sigma\sigma'}(\mathbf{q})$ are read off from the resulting matrix.
Written out component by component, Eq.~\eqref{eq:dyson} gives four coupled
equations:
\begin{align}
  K_{\uparrow\uparrow}(\mathbf{q}) &= \chi^0_\uparrow(\mathbf{q})
    + \chi^0_\uparrow(\mathbf{q}) U_{\uparrow\uparrow}
      K_{\uparrow\uparrow}(\mathbf{q})
    + \chi^0_\uparrow(\mathbf{q}) U_{\uparrow\downarrow}
      K_{\downarrow\uparrow}(\mathbf{q}),
    \notag\\
  K_{\downarrow\downarrow}(\mathbf{q}) &= \chi^0_\downarrow(\mathbf{q})
    + \chi^0_\downarrow(\mathbf{q}) U_{\downarrow\downarrow}
      K_{\downarrow\downarrow}(\mathbf{q})
    + \chi^0_\downarrow(\mathbf{q}) U_{\downarrow\uparrow}
      K_{\uparrow\downarrow}(\mathbf{q}),
    \notag\\
  K_{\uparrow\downarrow}(\mathbf{q}) &=
      \chi^0_\uparrow(\mathbf{q}) U_{\uparrow\uparrow}
      K_{\uparrow\downarrow}(\mathbf{q})
    + \chi^0_\uparrow(\mathbf{q}) U_{\uparrow\downarrow}
      K_{\downarrow\downarrow}(\mathbf{q}),
    \notag\\
  K_{\downarrow\uparrow}(\mathbf{q}) &=
      \chi^0_\downarrow(\mathbf{q}) U_{\downarrow\downarrow}
      K_{\downarrow\uparrow}(\mathbf{q})
    + \chi^0_\downarrow(\mathbf{q}) U_{\downarrow\uparrow}
      K_{\uparrow\uparrow}(\mathbf{q}),
  \label{eq:coupled}
\end{align}
which are solved independently at each $\mathbf{q}$.

Only the bare bubbles are computed in real space. They are Fourier transformed
site by site with Eq.~\eqref{eq:sitematrix}, and the Dyson equation is solved
at each $\mathbf{q}$ in spin and site space, with a local kernel
$U_{\sigma\sigma'}$ on each site that carries one; $\chi_{nn}$ and $\chi_{zz}$ follow by
projecting the result on the phase vector $\phi$ of Eq.~\eqref{eq:cellsum}.
For one site per cell this is Eq.~\eqref{eq:dyson}. Since the kernel is
defined per site, the same $U_{\sigma\sigma'}$ applies in every cell. In FeSe only the Fe sites carry a kernel; the Se sites enter $\chi^0$ but keep their bare response. For a local kernel the instability is set by the largest eigenvalue $\lambda_\mathrm{max}(\mathbf{q})$ of the per-spin site matrix $\chi^0_{\sigma,ij}(\mathbf{q})$ (Sec.~\ref{subsec:ft}), restricted to the sites that carry a kernel: in a nonmagnetic state with the same Stoner parameter $I$ [Eq.~\eqref{eq:stoner}] on these sites, the spin response diverges where $I\lambda_\mathrm{max}=1$. For one site per cell $\lambda_\mathrm{max}=\chi^0_\sigma(\mathbf{q})$, and $I\lambda_\mathrm{max}$ is the Stoner product $I\chi^0_\sigma(\mathbf{q})$.

The off-diagonal elements of $\mathbf{U}$ satisfy
$U_{\uparrow\downarrow}=U_{\downarrow\uparrow}$ in the absence of
spin-orbit coupling~\cite{Rajagopal1973,Gunnarsson1976LSD,Gunnarsson1977erratum}.
The diagonal elements are equal only in the non-spin-polarised case; in a
spin-polarised system they are evaluated at different spin densities and
are in general distinct. The combination that governs the spin channel is
the Stoner parameter
\begin{equation}
  I = \tfrac{1}{2}\left(U_{\uparrow\uparrow}+U_{\downarrow\downarrow}\right)
      - U_{\uparrow\downarrow},
  \label{eq:stoner}
\end{equation}
which in the nonmagnetic state, where
$U_{\uparrow\uparrow}=U_{\downarrow\downarrow}$, reduces to
$I=U_{\uparrow\uparrow}-U_{\uparrow\downarrow}$. It equals $\lim_{m\to0}B_\mathrm{xc}/m$, the
nonmagnetic limit of $I_\mathrm{xc}=B_\mathrm{xc}/m$ of
Ref.~\onlinecite{katsnelson2004magnetic} [their Eq.~(10)], and the
Stoner parameter of Ref.~\onlinecite{Gunnarsson1976}. The charge channel is
governed by $U_{\uparrow\uparrow}+U_{\uparrow\downarrow}$. In the scans we take $U_{\downarrow\downarrow}=U_{\uparrow\uparrow}$, which is exact in the nonmagnetic states and an approximation in the ordered ones (Sec.~\ref{sec:ordered}), and $U_{\uparrow\downarrow}=0.1\,U_{\uparrow\uparrow}$, so that $I=0.9\,U_{\uparrow\uparrow}$. The small off-diagonal element has a simple origin. By the spin-scaling relation the exchange energy is a sum of separate spin-up and spin-down terms~\cite{oliver1979spin}, so exchange contributes only to the diagonal elements, and $U_{\uparrow\downarrow}$ comes from correlation alone. The kernels calculated from Eq.~\eqref{eq:Ukernel} confirm this: $U_{\uparrow\downarrow}$ is 7--10\% of the diagonal elements in all systems studied here (Appendix~\ref{app:U}). In a nonmagnetic state the spin and charge channels decouple, so the spin response depends on $\mathbf{U}$ only through $I$, and the value of the ratio affects only the charge channel. The figure legends quote the elements $U_{\sigma\sigma'}$, and the proximity to an instability is quoted in terms of $I$.
The denominator $[\mathbf{1} -  \boldsymbol{\chi}^0(\mathbf{q})\,\mathbf{U}]$
plays the role of the Stoner screening: when
$I\cdot\chi^0_\sigma(\mathbf{q}^*)$ approaches unity at some wave vector
$\mathbf{q}^*$, the interacting spin susceptibility diverges, signaling the
onset of a broken-symmetry ground state at that ordering vector.

The charge and longitudinal spin susceptibilities then follow
from the response of the total density $n = n_\uparrow + n_\downarrow$
to a spin-independent (charge) perturbation and of
$S_z = \tfrac{1}{2}(n_\uparrow - n_\downarrow)$ to a spin-antisymmetric
(magnetic) one, respectively,
\begin{align}
  \chi_{nn}(\mathbf{q}) &=
    K_{\uparrow\uparrow} + K_{\downarrow\downarrow}
    + K_{\uparrow\downarrow} + K_{\downarrow\uparrow},
  \label{eq:chinn} \\
  \chi_{zz}(\mathbf{q}) &=
    \tfrac{1}{4}\!\left(K_{\uparrow\uparrow} + K_{\downarrow\downarrow}
    - K_{\uparrow\downarrow} - K_{\downarrow\uparrow}\right).
  \label{eq:chizz}
\end{align}

This assignment is fixed by the noninteracting limit: at
$\mathbf{U} = 0$ the off-diagonal components
$K_{\uparrow\downarrow} = K_{\downarrow\uparrow}$ vanish and
$K_{\sigma\sigma} \to \chi^0_\sigma$, so the charge channel reduces to
the full Lindhard result
$\chi_{nn}(\mathbf{q}) \to \chi^0_\uparrow(\mathbf{q}) +
\chi^0_\downarrow(\mathbf{q}) \equiv \chi^0_{nn}(\mathbf{q})$, while the
spin channel reduces to
$\chi_{zz}(\mathbf{q}) \to \tfrac{1}{4}\,\chi^0_{nn}(\mathbf{q})$.
All susceptibilities here are per unit cell, and for the bare quantities
$\chi^0_{zz}=\tfrac14\chi^0_{nn}=\tfrac12\chi^0_{\sigma}$, where
$\chi^0_{\sigma}$ is the single-spin bubble of Eq.~\eqref{eq:chi0}.

Equation~\eqref{eq:dyson}, like Eq.~(19) of Ref.~\onlinecite{katsnelson2004magnetic}, contains only the exchange-correlation part of the kernel. This is complete for the longitudinal spin response of a nonmagnetic state, where the Hartree term cancels, but not for the charge response, whose full kernel also contains the repulsive Coulomb term~\cite{Johannes2006}. The $\chi_{nn}$ obtained here is therefore the response to the external plus Hartree potential, enhanced by exchange and correlation; it shows how a local kernel reshapes $\chi^0_{nn}$, but its pole does not signal a CDW instability. For the CDW systems we base the discussion on the bare $\chi^0_{nn}(\mathbf{q})$.

\subsection{Fourier Transform and Spectral Analysis}
\label{subsec:ft}

The spin-resolved real-space bare bubbles $\chi^0_{\sigma,ij}(\mathbf{R})$ are
computed for all pairs (site $i$, site $j$, lattice translation $\mathbf{R}$)
with $|\mathbf{d}_{ij}| \le r_{\mathrm{cut}}$ and Fourier transformed to
reciprocal space by direct summation over the bond vectors,
\begin{equation}
  \chi^0_\sigma(\mathbf{q}) = \sum_{i,j,\mathbf{R}}
    \chi^0_{\sigma,ij}(\mathbf{R})\,
    e^{i\mathbf{q}\cdot\mathbf{d}_{ij}},
  \qquad
  \mathbf{d}_{ij} = \mathbf{r}_j + \mathbf{R} - \mathbf{r}_i,
  \label{eq:ft}
\end{equation}
where $\mathbf{d}_{ij}$ is the bond from site $i$ in the home cell to site $j$
in the cell at lattice translation $\mathbf{R}$. The positions $\mathbf{r}_i$
must be those used in the self-consistent calculation, which can differ from
the input positions by lattice vectors. We determine them from the requirement
that the bonds of every site form one complete sphere of radius
$r_\mathrm{cut}$.
{
Keeping the site indices and omitting the intracell displacements from the phase defines the site matrix,
\begin{equation}
\chi^{0}_{\sigma,ij}(\mathbf{q}) = \sum_{\mathbf{R}} \chi^{0}_{\sigma,ij}(\mathbf{R})\, e^{i\mathbf{q}\cdot\mathbf{R}},
\label{eq:sitematrix}
\end{equation}
which is Hermitian and periodic in $\mathbf{q}$. The cell-summed bubble of Eq.~(\ref{eq:ft}) is its expectation value in the phase vector $\phi_i = e^{i\mathbf{q}\cdot\mathbf{r}_i}$,
\begin{equation}
\chi^{0}_{\sigma}(\mathbf{q}) = \sum_{ij} \phi_i^{*}\, \chi^{0}_{\sigma,ij}(\mathbf{q})\, \phi_j ,
\label{eq:cellsum}
\end{equation}
so that solving the Dyson equation for $\chi^{0}_{\sigma}(\mathbf{q})$ alone is exact precisely when $\phi$ is an eigenvector of $\chi^{0}_{\sigma,ij}(\mathbf{q})$ (Sec.~\ref{subsec:dyson}).
}
For cells with more than one site the intracell displacement enters the
phase, and $\chi^0(\mathbf{q})$ is the response of the site charges to a
plane-wave perturbation. {It is even in $\mathbf{q}$, as required by time-reversal symmetry, but it is not periodic under
$\mathbf{q}\to\mathbf{q}+\mathbf{G}$: the shift multiplies the phase on site $i$ by $e^{i\mathbf{G}\cdot\mathbf{r}_i}$, which changes the relative phase of sites that are not related by lattice translations. A Lindhard function with constant matrix elements ignores these phases and is therefore periodic~\cite{Heil2014}.} A curve can therefore change slope where the path
turns at a zone-boundary point, and points in the neighbouring zone, marked
with primes, are not equivalent to their first-zone images. Because the site
charges enter without an atomic form factor, values far outside the first zone
are not directly comparable with plane-wave calculations. The instability
criterion for a site-local kernel is set by the eigenvalues of the site
matrix $\chi^0_{\sigma,ij}(\mathbf{q})$, which are periodic and do not depend
on these conventions.

The sum runs over every site pair in the cell, so $\chi^0_\sigma(\mathbf{q})$
is a per-unit-cell quantity.

The transformed bubbles, arranged as the site matrices $\chi^0_{\sigma,ij}(\mathbf{q})$ of Eq.~\eqref{eq:sitematrix}, enter the Dyson equation in spin and site space, which is solved independently at each $\mathbf{q}$ (Sec.~\ref{subsec:dyson}); for one site per cell this is Eq.~\eqref{eq:dyson} with the diagonal matrix of Eq.~\eqref{eq:chi0mat}. The interacting $\chi_{nn}(\mathbf{q})$ and
$\chi_{zz}(\mathbf{q})$ are therefore obtained directly on the chosen
high-symmetry paths, and no Fourier transform of the interacting response
itself is required.
At $\mathbf{q}=\mathbf{0}$ the bare charge susceptibility approximately satisfies the
compressibility sum rule
\begin{equation}
  \chi^0_{nn}(\mathbf{q}=\mathbf{0}) =
  \chi^0_\uparrow(\mathbf{0}) + \chi^0_\downarrow(\mathbf{0}) =
  N(\varepsilon_F),
  \label{eq:sumrule}
\end{equation}
where $N(\varepsilon_F)$ is the density of states at the Fermi level, per
unit cell, of the orbitals retained in the projection window, thermally
averaged at finite $T$. The equality is exact only for a complete orthonormal set of local orbitals; with the muffin-tin-head projection used here the two agree to within $13\%$ (Supplemental Material). The same limit serves
as an accuracy test in Ref.~\onlinecite{Heil2014}; the two quantities are
compared for each system in the Supplemental Material.

\subsection{Computational Details}
\label{subsec:compdet}

All calculations use the full-potential linear muffin-tin orbital code RSPt~\cite{Wills2010}, which makes no shape approximation to the density or the potential, with the PBE exchange-correlation functional. The structures are listed in Table~\ref{tab:struct}. The calculations are non-spin-polarised except in Sec.~\ref{sec:ordered}, where they are collinear and spin-polarised. Spin-orbit coupling is not included. For bcc~Cr the non-spin-polarised calculations use the experimental lattice constant (Table~\ref{tab:struct}). At that volume PBE gives a sublattice moment of $1.18\,\mu_B$ in the G-type antiferromagnetic state, above the measured $0.62\,\mu_B$~\cite{Fawcett1988}; the magnetically ordered calculations of Sec.~\ref{sec:ordered} therefore use the volume that reproduces the measured moment, $74.0$~a.u.$^3$ per atom ($a=5.29$~a.u.), which gives an atomic moment of $0.62\,\mu_B$. 
The Green's functions entering Eq.~\eqref{eq:chi0} are taken from self-consistent calculations. Brillouin-zone integrations use Fermi-Dirac smearing of width $W=0.005$~Ry, which also sets the temperature in Eq.~\eqref{eq:chi0}, $T=W/k_B\approx790$~K. For bcc~Fe, $W$ is varied (Sec.~\ref{subsec:dw}). We use 1200 Matsubara frequencies for all systems. The $k$ meshes (Table~\ref{tab:comp}) cover the full zone without symmetry reduction, as the intersite Green's functions require. The intersite bubbles are computed for all pairs with $|\mathbf{d}_{ij}|\le r_\mathrm{cut}$. For LaTe$_3$, reducing $r_\mathrm{cut}$ from $83.0$ to $41.5$~a.u.\ changes $\chi^0_{nn}$ by $1.2\%$ on average and at most $6\%$ along the path, and moves its maximum by less than $2\%$ of the $\Gamma$-$Z$ distance. 
The kernel $\mathbf{U}$ is scanned over the values given with each figure; the values calculated from Eq.~\eqref{eq:Ukernel} are listed in Appendix~\ref{app:U}. The scanned values bracket the calculated kernel for NbSe$_2$; for LaTe$_3$ they exceed it, to make the enhancement visible, and for Cr and FeSe they stay below it, because the calculated kernels place the nonmagnetic states of both past the pole 
($I\lambda_\mathrm{max}\approx1.26$ for Cr and about $1.75$ for FeSe, with the $I$ of the antiferromagnetic state of each, Appendix~\ref{app:U}), consistent with the magnetic ground states that density functional theory gives for both. With the calculated kernels the CDW compounds stay far from an instability, with $I\lambda_\mathrm{max}=0.52$ for NbSe$_2$ and $0.16$ for LaTe$_3$. Wave vectors are given in Cartesian units of $2\pi/a$, $2\pi/b$ and
$2\pi/c$, except for hexagonal NbSe$_2$, where they are given in units of the
reciprocal lattice vectors.

\begin{table}[h]
\caption{Crystal structures used in this work. The magnetic cells of
Sec.~\ref{sec:ordered} keep the same lattice parameters, except for bcc~Cr
(Sec.~\ref{subsec:compdet}). $N_\mathrm{at}$ is the number of atoms in the
cell; the sites that enter $\chi^0$ are given in Table~\ref{tab:comp}.
Internal coordinates are listed in the Supplemental Material; the FeSe structure is from Ref.~\onlinecite{Kumar2010} and the bcc~Fe lattice parameter from Ref.~\onlinecite{Sandratskii1992}.}
\label{tab:struct}
\begin{ruledtabular}
\begin{tabular}{llcccc}
System & Space group & $a$ (\AA) & $b$ (\AA) & $c$ (\AA) & $N_\mathrm{at}$ \\
\hline
bcc Fe      & $Im\bar{3}m$ (229) & 2.789 & --     & --     & 1 \\
bcc Cr      & $Im\bar{3}m$ (229) & 2.884 & --     & --     & 1 \\
FeSe        & $P4/nmm$ (129)     & 3.768 & --     & 5.519  & 4 \\
2H-NbSe$_2$ & $P6_3/mmc$ (194)   & 3.440 & --     & 12.550 & 6 \\
LaTe$_3$    & $Cmcm$ (63)        & 4.392 & 26.247 & 4.417  & 8 \\
\end{tabular}
\end{ruledtabular}
\end{table}

Cr and Fe are body-centred cubic. FeSe is tetragonal: the Fe atoms form a square planar
lattice, each Fe is tetrahedrally coordinated by Se atoms lying alternately
above and below the Fe plane, and the FeSe layers are stacked along $c$ with
no atoms between them. 2H-NbSe$_2$ is hexagonal; within each NbSe$_2$ layer
the Nb atom is in trigonal-prismatic coordination by Se, and the 2H stacking
places two layers in the unit cell. LaTe$_3$ is orthorhombic and is described
in Sec.~\ref{subsec:late3}. For LaTe$_3$ we use the primitive cell, which
contains two formula units. The coordinates of the high-symmetry points are
given in the figure captions.

\begin{table}[h] 
\caption{Computational parameters. The $k$ meshes are shifted by half a mesh step, except for 2H-NbSe$_2$, where a $\Gamma$-centred mesh preserves the threefold symmetry. $r_\mathrm{cut}$ is the real-space cutoff of the intersite bubbles. $\chi^0_{nn}(\Gamma)$ is per unit cell; it approximates the thermally averaged density of states of the retained orbitals [Eq.~\eqref{eq:sumrule}]; the two are compared in the Supplemental Material.
The retained orbitals carry most of $N(\varepsilon_F)$. In 2H-NbSe$_2$ the Se-derived Fermi-surface pocket holds about 5\% of $N(\varepsilon_F)$ and contributes almost nothing along $\Gamma$--$M$~\cite{Johannes2006}, so only Nb~$4d$ is kept; in FeSe the Fe--Se bonds carry a sizeable part of the intersite response (Supplemental Material), so Se~$4p$ is kept.
} 
\label{tab:comp} 
\begin{ruledtabular} 
\begin{tabular}{lcccc} 
System & $k$ mesh & $r_\mathrm{cut}$ (a.u.) & Orbitals & $\chi^0_{nn}(\Gamma)$ (Ry$^{-1}$) \\ \hline 2H-NbSe$_2$ & $48\times48\times12$ & 32.5 & Nb~$4d$ & 28.0 \\ LaTe$_3$ & $30\times30\times30$ & 83.0 & Te~$5p$ (Te sites) & 22.5 \\ bcc Cr & $32\times32\times32$ & 33.0 & Cr~$4s$, $4p$, $3d$ & 9.26 \\ FeSe & $36\times36\times24$ & 35.6 & Fe~$3d$, Se~$4p$ & 59.0 \\ bcc Fe & $36\times36\times36$ & 36.9 & Fe~$3d$ & see Fig.~\ref{fig:stoner} \\ 
\end{tabular} 
\end{ruledtabular} 
\end{table}

\subsection{Density-Wave Instabilities and Physical Interpretation; the Test Case of bcc Fe}
\label{subsec:dw}

The different responses discussed in the previous sections correspond to different types of density waves. For instance, in a CDW system, the charge density acquires a spatial modulation at
wave vector $\mathbf{Q}_\mathrm{CDW}$, typically accompanied by a
periodic lattice distortion (Peierls distortion), while the spin
density remains spatially uniform
\cite{Gruner1988,Peierls1955}.
  A peak in the bare charge susceptibility $\chi^0_{nn}(\mathbf{q})$ marks the wave vectors at which the electrons respond most strongly to a charge modulation, such as that of a periodic lattice distortion; whether a CDW forms also depends on the electron-phonon coupling~\cite{Johannes2006,Weber2011}.

\begin{figure*}[htbp]
  \centering
  \includegraphics[width=\linewidth]{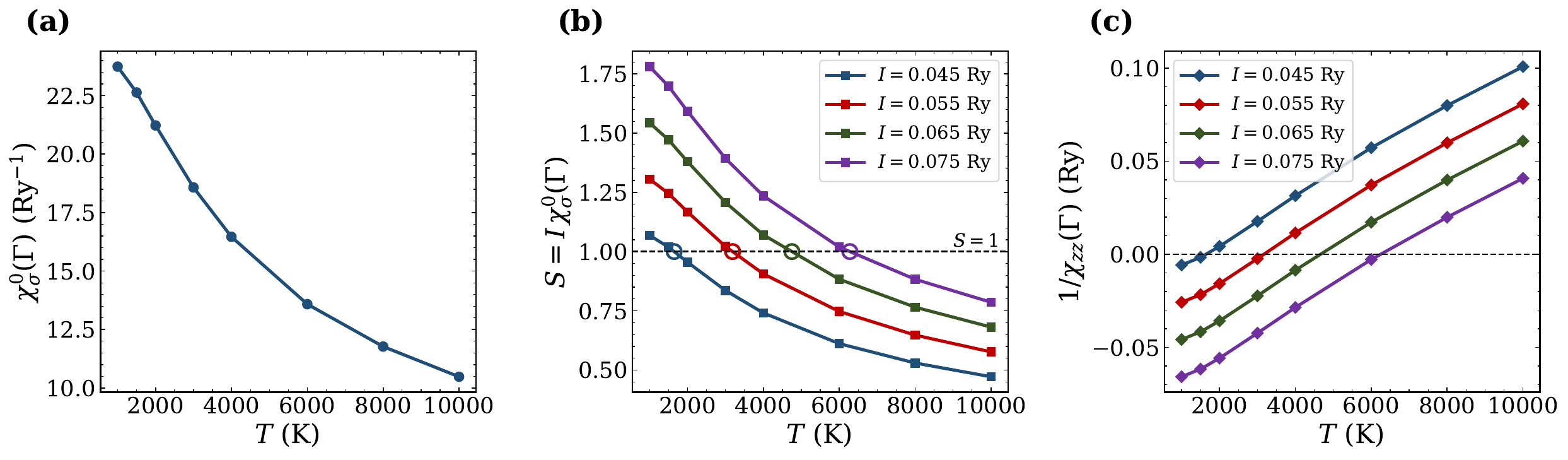}
  \caption{Test of the RPA Dyson solver against the ferromagnetic Stoner
    instability of bcc~Fe, evaluated at $\mathbf{q}=\Gamma$ as a function of
    the electronic temperature $T$.
    (a) Bare per-spin susceptibility $\chi^0_\sigma(\Gamma)$, which decreases
    monotonically with $T$.
    (b) Stoner product $S = I\chi^0_\sigma(\Gamma)$ for four values of the
    Stoner parameter $I$ [Eq.~\eqref{eq:stoner}], here $U_{\uparrow\uparrow}-U_{\uparrow\downarrow}$ since the state is nonmagnetic; the
    dashed line marks the critical value $S=1$ and the open circles the
    resulting mean-field Curie points.
    (c) Inverse of the interacting longitudinal spin susceptibility $1/\chi_{zz}(\Gamma)$, obtained from the interacting response matrix $\mathbf{K}$ of Eq.~\eqref{eq:dyson} through Eq.~\eqref{eq:chizz}, which
    changes sign at the same temperature.
        For each $I$ the three panels cross their respective thresholds at the same
    temperature, demonstrating internal consistency of the solver. The
    resulting mean-field transition temperatures exceed the experimental Curie
    point of bcc~Fe, as expected for Stoner theory, which omits the fluctuations of local
    moments~\cite{Gunnarsson1976}.}
  \label{fig:stoner}
\end{figure*}

In SDW systems the situation is reversed: the spin density is
periodically modulated (equivalent to two out-of-phase CDWs for
spin-up and spin-down electrons), while the charge density remains
nearly uniform \cite{Overhauser1962}.
A peak in the longitudinal spin susceptibility $\chi_{zz}(\mathbf{q})$
at the SDW ordering vector $\mathbf{Q}$ is the usual indicator of this
instability.
The Stoner criterion, $\tfrac12 I\chi^0_{nn}(\mathbf{Q})=I\chi^0_\sigma(\mathbf{Q})\geq 1$,
with $\chi^0_\sigma=\tfrac12\chi^0_{nn}$ the per-spin bare susceptibility, 
marks the
pole of the nonmagnetic spin response, 
and determines whether the RPA enhancement is sufficient
to actually drive the system magnetic; the closeness of the product
$I \cdot \chi^0_\sigma(\mathbf{Q})$ to unity is a measure of the
proximity to the instability.

Figure~\ref{fig:stoner} tests the Dyson solver developed here on the ferromagnetic
instability of bcc~Fe at $\mathbf{q}=\Gamma$: the numerical solution of
Eq.~\eqref{eq:dyson} must diverge exactly where $I\chi^0_\sigma(\Gamma)=1$, and the
temperature dependence of $\chi^0_\sigma(\Gamma)$ must give ordering temperatures close to
those obtained from Stoner theory. As the temperature increases,
$\chi^0_\sigma(\Gamma)$ decreases [Fig.~\ref{fig:stoner}(a)], because thermal
broadening lowers the effective density of states at $\varepsilon_F$. For each
Stoner parameter $I$, the product $I\chi^0_\sigma(\Gamma)$ passes through 1
[Fig.~\ref{fig:stoner}(b)] at the same temperature at which
$1/\chi_{zz}(\Gamma)$ changes sign [Fig.~\ref{fig:stoner}(c)], as the Stoner
criterion requires. The crossing moves from about 1700~K at $I=0.045$~Ry to
about 6300~K at $I=0.075$~Ry. The Stoner parameter of Fe in the local
spin-density approximation, $I\approx0.068$~Ry~\cite{Gunnarsson1976,Janak1977},
lies between the curves for $0.065$ and $0.075$~Ry, which cross at about 4800
and 6300~K and represent ordering temperatures from Stoner theory. This should be compared to results of Gunnarsson~\cite{Gunnarsson1976} who obtained 4400 to 6200~K with the same Stoner
model, and the similarity between earlier works on the present implementation represents a numerical test that falls out favorably. The values obtained here should be compared to an experimental ordering temperature of $T_C$ of 1043~K, that are in decent agreement with theory that takes into account thermal fluctuations of the directions of the magnetic moments~\cite{LKAG1987}.

\section{Results and Discussion}
\label{sec:results}

\begin{figure*}[htbp]
  \centering
  \includegraphics[width=\textwidth]{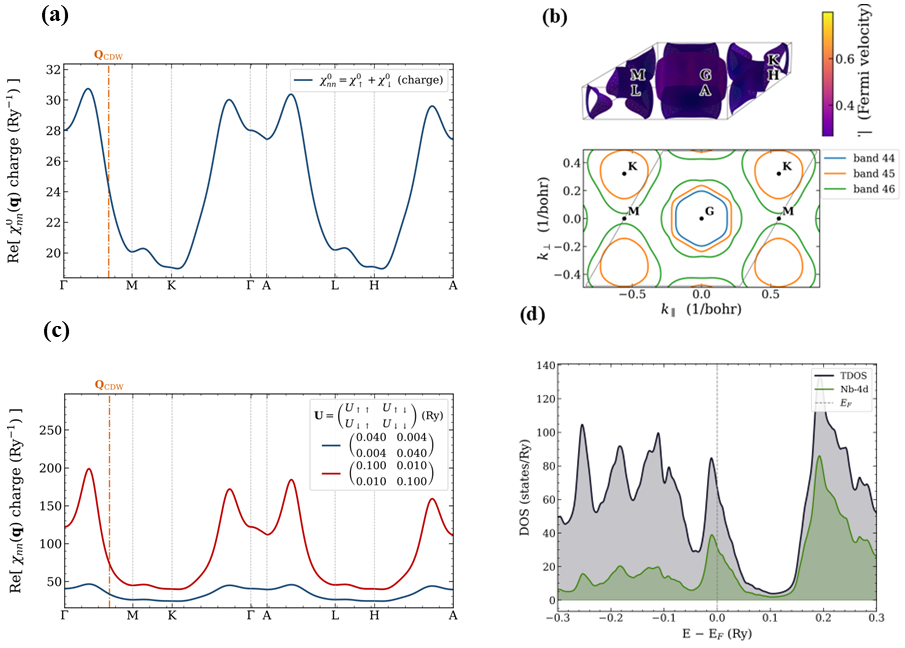}
  \caption{Results for 2H-NbSe$_2$.
        (a) Bare charge susceptibility
    $\chi^0_{nn}(\mathbf{q})=\chi^0_\uparrow+\chi^0_\downarrow$ along
    $\Gamma$-$M$-$K$-$\Gamma$-$A$-$L$-$H$-$A$, with $M=(\tfrac12,0,0)$,
    $K=(\tfrac13,\tfrac13,0)$, $A=(0,0,\tfrac12)$, $L=(\tfrac12,0,\tfrac12)$
    and $H=(\tfrac13,\tfrac13,\tfrac12)$ in reciprocal-lattice units; the
    vertical line marks $\mathbf{Q}_\mathrm{CDW}$.
    (b) Calculated Fermi surface in the hexagonal Brillouin zone: (top)
    three-dimensional sheets coloured by the magnitude of the Fermi
    velocity $|\nabla_\mathbf{k}E|$, and (bottom) a band-resolved
    cross-section, showing the two concentric quasi-two-dimensional
    Nb~$4d$ barrels around $\Gamma$ and around the $K$/$H$ zone edges
    together with the small Se-derived pocket at the zone centre.
   (c) Interacting charge susceptibility $\chi_{nn}(\mathbf{q})$ from the RPA Dyson equation with the exchange-correlation kernel only (Sec.~\ref{subsec:dyson}), for two interaction strengths
    $(U_{\uparrow\uparrow},U_{\uparrow\downarrow}$, in Ry).
    (d) Total density of states and the Nb~$4d$ projection summed over both Nb
sites; $\varepsilon_F$ (dashed) lies on a Nb~$4d$ peak. The remaining weight
at $\varepsilon_F$ is Se~$4p$ and interstitial. The projections are
muffin-tin quantities and are not expected to add up to the total. All results are for a thermal smearing width $W = 0.005$~Ry ($T \approx 790$~K).}
  \label{fig:nbse2}
\end{figure*}
We now turn our attention to four benchmark systems: two CDW materials (2H-NbSe$_2$
and LaTe$_3$), the SDW material bcc~Cr, and FeSe, which exhibits strong stripe
spin fluctuations.
For each system we show (i) the electronic density of states with its
orbital projection, (ii) the Fermi surface, (iii) the bare
susceptibility $\chi^0_{nn}(\mathbf{q})$ along high-symmetry directions
of the Brillouin zone, and (iv) the interacting susceptibility obtained
from the RPA Dyson equation (the charge channel $\chi_{nn}(\mathbf{q})$
for the CDW systems and the longitudinal spin channel
$\chi_{zz}(\mathbf{q})$ for the SDW systems) for a scan of the on-site
interaction strength characterised by the $U$ matrix [Eq.~\eqref{eq:Ukernel}].
Together these representations connect the electronic structure and
Fermi surface geometry of each material to the wave-vector and strength
of its incipient density-wave instability.

\section{Charge Density Wave Systems}
\label{sec:cdw}

\subsection{2H-NbSe$_2$}
\label{subsec:nbse2}

2H-NbSe$_2$ is hexagonal ($P6_3/mmc$, Table~\ref{tab:struct}). Below
$T_\mathrm{CDW}\approx33$~K it develops a nearly commensurate $3\times3$ CDW
with $\mathbf{Q}_\mathrm{CDW}\approx(\tfrac13,0,0)$, two thirds of the way from 
$\Gamma$ to $M$, and it superconducts below
$T_c\approx7.2$~K~\cite{Wilson1975,Kiss2007}. First-principles calculations
find no Fermi-surface nesting at $\mathbf{Q}_\mathrm{CDW}$~\cite{Johannes2006},
and inelastic x-ray scattering shows that the CDW comes from a strongly
momentum-dependent electron-phonon coupling, which softens an acoustic phonon
over an extended range of $\mathbf{q}$ around
$\mathbf{Q}_\mathrm{CDW}$~\cite{Weber2011}.

At $\varepsilon_F$, which lies on a DOS peak, the largest projected
contribution to the density of states [Fig.~\ref{fig:nbse2}(d)] comes from
Nb~$4d$ states. The Fermi surface of 
Fig.~\ref{fig:nbse2}(b) has a known shape: two Nb~$4d$ cylinders around
$\Gamma$, two around the $K$--$H$ edges and a small Se-derived pocket at
$\Gamma$~\cite{Rossnagel2001,Johannes2006}.  

The bare susceptibility [Fig.~\ref{fig:nbse2}(a)] varies only between about
$19$ and $31~\mathrm{Ry}^{-1}$ along the path. It has four maxima of nearly
equal height, about $30~\mathrm{Ry}^{-1}$, on $\Gamma$--$M$, $K$--$\Gamma$,
$A$--$L$ and $H$--$A$. The highest lies on $\Gamma$--$M$, the direction of the
CDW, at $\mathbf{q}\approx(0.18,0,0)$, about a third of the way from $\Gamma$ to $M$ and roughly half of
$\mathbf{Q}_\mathrm{CDW}$. Beyond it the
response falls only slowly: at $\mathbf{Q}_\mathrm{CDW}$,
$\chi^0_{nn}\approx24~\mathrm{Ry}^{-1}$, still about $80\%$ of the maximum, so
$\mathbf{Q}_\mathrm{CDW}$ lies on the flank of a broad maximum rather than at a
peak. This agrees with the picture that Weber \emph{et al.} ~\cite{Weber2011} drew from the
phonon dispersion: the acoustic phonon softens over an extended range of
wave vectors along $\Gamma$--$M$, the electronic states supply a broadly
enhanced response, and the momentum dependence of the electron-phonon
coupling selects $\mathbf{Q}_\mathrm{CDW}$ within it~\cite{Weber2011}.
Johannes \emph{et al.} likewise found only a weak, broad maximum of
$\mathrm{Re}\,\chi^0$ near $\mathbf{Q}_\mathrm{CDW}$, at $(0.31,0,0)$, and no
nesting feature in $\mathrm{Im}\,\chi^0$ at
$\mathbf{Q}_\mathrm{CDW}$~\cite{Johannes2006}. Their maximum lies closer to
$\mathbf{Q}_\mathrm{CDW}$ than ours. They set the matrix elements to unity and
keep only intraband terms, whereas we retain the Nb~$4d$ matrix elements; we
have not established which difference moves the maximum. Their band
decomposition shows almost no contribution from the Se-derived pocket along
$\Gamma$--$M$~\cite{Johannes2006}.

Solving Eq.~\eqref{eq:coupled} for the nonmagnetic case,
$\chi^0_\uparrow=\chi^0_\downarrow=\tfrac12\chi^0_{nn}$ and
$U_{\uparrow\uparrow}=U_{\downarrow\downarrow}$, the charge channel of
Eq.~\eqref{eq:chinn} becomes
\begin{equation}
  \chi_{nn}(\mathbf{q}) = \frac{\chi^0_{nn}(\mathbf{q})}
  {1 - \tfrac{1}{2}\,\chi^0_{nn}(\mathbf{q})
    \left(U_{\uparrow\uparrow}+U_{\uparrow\downarrow}\right)} .
  \label{eq:nbse2_rpa}
\end{equation}
Here $\chi^0_{nn}$ and $\chi_{nn}$ are per Nb site, half of the per-cell values plotted in Fig.~\ref{fig:nbse2}; the equation treats the two Nb sites of the cell as independent, whereas the curves in Fig.~\ref{fig:nbse2}(c) come from the full site-resolved Dyson equation (Sec.~\ref{subsec:dyson}). Below the pole, $\chi_{nn}$ increases monotonically with $\chi^0_{nn}$, so the
kernel raises the existing maxima but cannot move them. With
$U_{\uparrow\uparrow}=0.10$~Ry and $U_{\uparrow\downarrow}=0.010$~Ry per Nb
[Fig.~\ref{fig:nbse2}(c)], the highest maximum grows to about
$200~\mathrm{Ry}^{-1}$ and the minima to about $40~\mathrm{Ry}^{-1}$. These
maxima do not change $\mathbf{q}$, as compared to results of unenhanced susceptibilities, so that $\mathbf{Q}_\mathrm{CDW}$ remains on a flank with the theoretical maximum closer to the zone centre than the experimental value. The kernel also sharpens the contrast between the two vectors: at $\mathbf{Q}_\mathrm{CDW}$, $\chi_{nn}\approx73~\mathrm{Ry}^{-1}$, about $37\%$ of the maximum, compared with about $80\%$ for $\chi^0_{nn}$. A local electronic kernel therefore favours $\mathbf{Q}_\mathrm{CDW}$ even less than the bare response does, and its selection must come from the momentum dependence of the electron-phonon coupling~\cite{Weber2011}.

\subsection{LaTe$_3$}
\label{subsec:late3}

LaTe$_3$ crystallises in the orthorhombic space group $Cmcm$ (No.~63). As
usual for the rare-earth tritellurides, $b$ is the long stacking axis and the
two in-plane axes are nearly equal, $a\approx c$ (Table~\ref{tab:struct}).
Corrugated LaTe double layers alternate along $b$ with pairs of planar square
Te nets, and neighbouring nets are held together only by weak van der Waals
bonding~\cite{Gweon1998,Hong2022}. Of the three inequivalent Te sites, Te1
belongs to the LaTe layer and Te2 and Te3 form the structural nets. Each Te atom of such a net carries
about half an extra electron~\cite{Gweon1998}, so that the Te~$5p$ bands of the nets
are partly filled and metallic, with a small La~$5d$ admixture at
$\varepsilon_F$~\cite{Hong2022}. The CDW forms from states located in the nets. It is
unidirectional and incommensurate, with a period of about $3.5$ lattice
constants along $c$
[$\mathbf{Q}_\mathrm{CDW}\approx\tfrac27\,(2\pi/c)$]
\cite{Ru2008,Hong2022}, its observed transition temperature exceeds
450~K~\cite{Ru2008}, and large parts of the Fermi surface stay
ungapped below it~\cite{Gweon1998,Brouet2008}. The direction of the CDW is set by a small in-plane anisotropy. A glide plane makes the structure weakly orthorhombic, with $c$ slightly longer than $a$ (Table~\ref{tab:struct})~\cite{Straquadine2022,GalloFrantz2024}, and in other rare-earth tritellurides stress that raises $a/c$ turns the CDW from $c$ to $a$~\cite{Straquadine2022,GalloFrantz2024,Singh2024,Singh2023thesis}. The electron-phonon coupling that drives the CDW does not choose the direction by itself: the phonons coupled to it soften along both in-plane axes at the transition~\cite{Maschek2018,Singh2024}, and the weak orthorhombicity acts as a small field that favours $c$~\cite{Yao2006,Singh2024}. Only one orientation forms, because the two compete. In a model of the Te-net bands, a CDW along one direction gaps the Fermi surface that the other would need, and a single CDW is favoured when the transition temperature is high~\cite{Yao2006}; in LaTe$_3$, a CDW along $a$ appears for a short time when light weakens the one along $c$~\cite{Kogar2020}. First-principles calculations
attribute the choice of $\mathbf{Q}_\mathrm{CDW}$ to momentum-dependent
electron-phonon coupling, and not to nesting~\cite{Hong2022}.

\begin{figure*}[htbp]
  \centering
  \includegraphics[width=\textwidth]{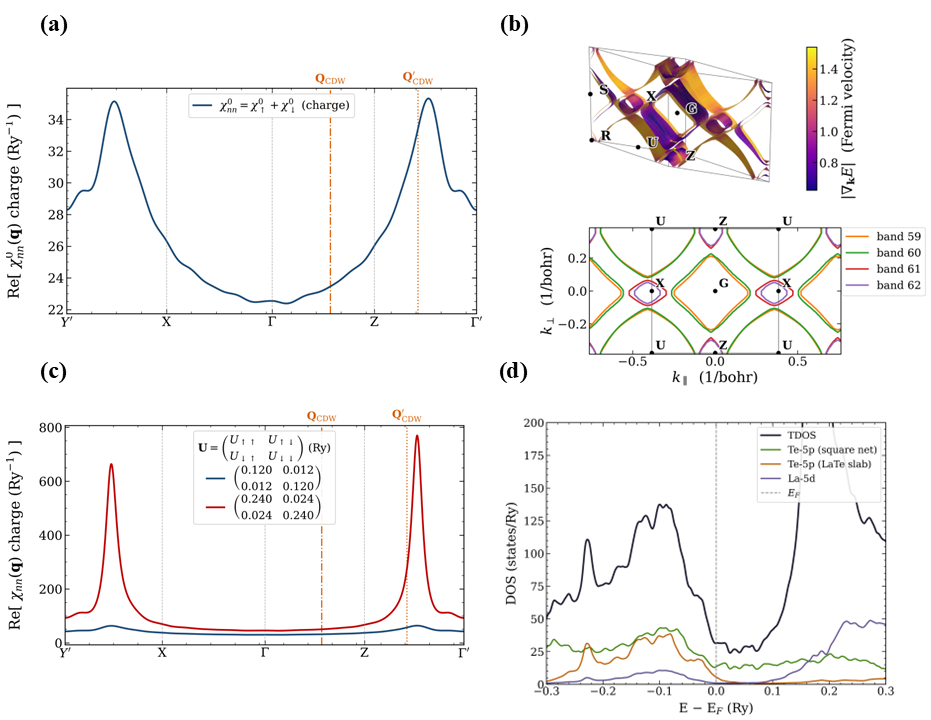}
  \caption{Results for LaTe$_3$.
   (a) Bare charge susceptibility $\chi^0_{nn}(\mathbf{q})$ of the Te sites
    along $Y'$-$X$-$\Gamma$-$Z$-$\Gamma'$: along the in-plane $a$ axis from
    $Y'=(2\pi/a,0,0)$ through the zone boundary $X=(1.03\,\pi/a,0,0)$ to
    $\Gamma$, and along the in-plane $c$ axis from $\Gamma$ through
    $Z=(0,0,\pi/c)$ to $\Gamma'=(0,0,2\pi/c)$; $Y'$ and $\Gamma'$ lie in the
    neighbouring Brillouin zone (Sec.~\ref{subsec:ft}). Vertical lines: the
    measured $\mathbf{Q}_\mathrm{CDW}$~\cite{Hong2022} and its mirror image about $Z$,
    {$\mathbf{Q}'_\mathrm{CDW}=\Gamma'-\mathbf{Q}_\mathrm{CDW}$, which gives the same
    modulation period but a different relative phase on the two Te atoms of a net (see text).}
    (b) Calculated Fermi surface: (top) three-dimensional sheets coloured
    by the magnitude of the Fermi velocity $|\nabla_\mathbf{k}E|$, and
    (bottom) a band-resolved cross-section, displaying the
    characteristic nearly-planar, weakly-warped Te~$5p$ sheets of the
    $R$Te$_3$ family; extended approximately parallel segments provide
    the geometric basis for nesting.
    (c) Interacting charge susceptibility $\chi_{nn}(\mathbf{q})$ from the RPA Dyson equation with the exchange-correlation kernel only (Sec.~\ref{subsec:dyson}), at $U_{\uparrow\uparrow}=0.12$ and $0.24$~Ry per Te, along the same path.
    (d) Total density of states with the La~$5d$ projection and the Te~$5p$
    projections of the square nets and of the LaTe layer; $\varepsilon_F$ lies in a
    shallow minimum of the Te~$5p$ states. The sharp peak about $0.18$~Ry above
    $\varepsilon_F$ comes from the empty La~$4f$ states, which are not among the
    projections, and is clipped. All results are for a thermal smearing width $W=0.005$~Ry
    ($T\approx790$~K).}
  \label{fig:late3}
\end{figure*}

The here calculated Fermi level lies in a shallow minimum of the Te~$5p$ density of states
[Fig.~\ref{fig:late3}(d)]. The narrow peak about $0.18$~Ry above
$\varepsilon_F$ comes from the empty La~$4f$ states. The Fermi surface
[Fig.~\ref{fig:late3}(b)] consists of nearly flat, weakly warped sheets formed
by the in-plane Te~$5p_x$ and $5p_z$ states of the structural nets~\cite{Brouet2004,Gweon1998}.

We compute the response of the six Te sites, which carry the states at
$\varepsilon_F$. The bare susceptibility [Fig.~\ref{fig:late3}(a)] comes from
the Te nets; the Te of the LaTe layer contributes little. Along
$\Gamma$--$Z$, the direction of the CDW, it rises from $22.5~\mathrm{Ry}^{-1}$
at $\Gamma$ to $23.5~\mathrm{Ry}^{-1}$ at $\mathbf{Q}_\mathrm{CDW}$, which lies
four sevenths of the way to $Z$, and keeps rising beyond $Z$ to a conspicuous
maximum of $35.3~\mathrm{Ry}^{-1}$. Along $\Gamma$--$X$ it behaves almost
identically, as expected for $a\approx c$, so the bare response does not select the direction of the CDW. First-principles calculations find the CDW along $c$ lower in energy than along $a$, and, in a monolayer, a much larger phonon linewidth at $\mathbf{Q}_\mathrm{CDW}$ along $c$ than at the corresponding wave vector along $a$~\cite{Hong2022}.

{The maximum lies close to the second marked vector, $\mathbf{Q}'_\mathrm{CDW}=\Gamma'-\mathbf{Q}_\mathrm{CDW}$, the mirror image of $\mathbf{Q}_\mathrm{CDW}$ about $Z$, where the response is $33.4~\mathrm{Ry}^{-1}$; the maximum lies slightly beyond it, by about a tenth of the $\Gamma$--$Z$ distance. The two vectors are not equivalent for the quantity plotted here. Time-reversal symmetry makes $\chi^0_{nn}(\mathbf{q})$ even, $\chi^0_{nn}(-\mathbf{q})=\chi^0_{nn}(\mathbf{q})$, but the plane-wave response of Eq.~\eqref{eq:ft} is not periodic in a cell with several sites (Sec.~\ref{subsec:ft}). $\mathbf{Q}'_\mathrm{CDW}$ differs from $-\mathbf{Q}_\mathrm{CDW}$ by the reciprocal lattice vector $\Gamma'$, and this shift multiplies the phase on site $i$ by $e^{i\Gamma'\cdot\mathbf{r}_i}$. The two vectors therefore give the same modulation period of the crystal~\cite{Ru2008}, but a different relative phase on the two Te atoms of a net. A Lindhard function with constant matrix elements ignores these phases and takes the same value at both vectors~\cite{Heil2014}. The same effect is well known in neutron scattering: the dynamical structure factor of a crystal with several atoms in the cell contains the phases $e^{i\mathbf{Q}\cdot\mathbf{r}_i}$ of the atoms, so a mode can be strong at $\mathbf{q}+\mathbf{G}$ and weak at $\mathbf{q}$~\cite{Lovesey1984}. In thin magnetic films, for example, the optical spin-wave branches are suppressed in the first Brillouin zone and dominate outside it~\cite{Taroni2011}. The difference reflects the smaller lattice of the Te net: for the electrons of a net, $\mathbf{Q}_\mathrm{CDW}$ and $\mathbf{Q}'_\mathrm{CDW}$ are distinct wave vectors, which are folded onto each other only by the weak potential of the LaTe layers~\cite{Brouet2004,Brouet2008}, and ARPES shows that within a single net the nested Fermi-surface sheets are connected by $\mathbf{Q}'_\mathrm{CDW}$~\cite{Brouet2008}. For this reason the CDW of the rare-earth tritellurides is described in the literature both by $\mathbf{Q}_\mathrm{CDW}\approx\tfrac27\,(2\pi/c)$, the value given by diffraction~\cite{Ru2008}, and by $\mathbf{Q}'_\mathrm{CDW}\approx\tfrac57\,(2\pi/c)$, the vector that nests the Fermi surface of a single net~\cite{Yumigeta2021}. The two describe the same lattice modulation. Our calculation, which keeps the phases of the Te sites, distinguishes them: the electronic response of the nets peaks near $\mathbf{Q}'_\mathrm{CDW}$, not near $\mathbf{Q}_\mathrm{CDW}$. The eigenvalues of the site matrix [Eq.~\eqref{eq:sitematrix}], which set the instability for a site-local kernel, are periodic; seen from the first zone, the largest one peaks just below $\mathbf{Q}_\mathrm{CDW}$, with the same offset. The offset hardly changes when the real-space cutoff is halved.} The Lindhard function of Hong \emph{et al.}, which has no site phases and is therefore the same at both vectors, peaks away from $\mathbf{Q}_\mathrm{CDW}$~\cite{Hong2022}. Our maximum lies close to $\mathbf{Q}'_\mathrm{CDW}$; the small remaining offset is consistent with the momentum-dependent electron-phonon coupling that Hong \emph{et al.} find to drive the CDW~\cite{Hong2022}.
The kernel enhances the charge response without changing its shape
[Fig.~\ref{fig:late3}(c)]: at $U_{\uparrow\uparrow}=0.12$ and $0.24$~Ry per Te the
maximum grows to about $64$ and $770~\mathrm{Ry}^{-1}$, and it stays where it
was.

The two CDW compounds therefore differ. In LaTe$_3$ the bare response has one marked maximum close to the observed modulation, at its image $\mathbf{Q}'_\mathrm{CDW}$, so the electronic structure of the Te nets already comes close to selecting the CDW vector. In NbSe$_2$ the response is weakly modulated, with $\mathbf{Q}_\mathrm{CDW}$ on the flank of the largest of several similar maxima, and the selection of $\mathbf{Q}_\mathrm{CDW}$ requires the momentum dependence of the electron-phonon coupling~\cite{Weber2011}, which a local electronic kernel does not contain.

\section{Spin Density Wave Systems}
\label{sec:sdw}

\subsection{bcc Cr}
\label{subsec:cr}

Chromium is body-centred cubic ($Im\bar3m$, Table~\ref{tab:struct}) and is the
standard itinerant spin-density-wave antiferromagnet~\cite{Fawcett1988}. Below
$T_N=311$~K it orders with $\mathbf{Q}_\mathrm{SDW}=(1-\delta)H$ along a cube axis,
where $H=(0,0,1)\,2\pi/a$ and $\delta\approx0.048$ at low temperature, so the
modulation period is
about 21 lattice constants~\cite{Fawcett1988}. The order is usually attributed
to nesting between the electron `jack' centred at $\Gamma$ and the hole
octahedron centred at $H$; $\delta$ is nonzero because the two sheets differ
slightly in size~\cite{Lomer1962,Fawcett1988}.

\begin{figure*}[htbp]
  \centering
  \includegraphics[width=\textwidth]{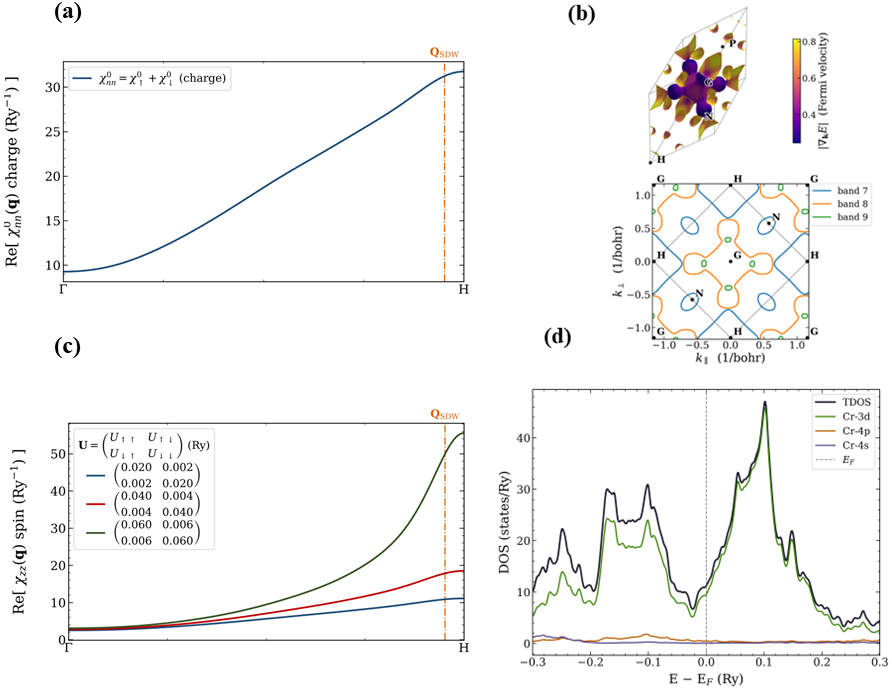}
  \caption{Results for bcc~Cr.
    (a) Bare susceptibility $\chi^0_{nn}(\mathbf{q})$ along
    $\Gamma$-$H$, with $H=(0,0,1)\,2\pi/a$, whose $\mathbf{q}$-dependence
    coincides with that of the bare spin channel
    ($\chi^0_{zz}=\chi^0_{nn}/4$); the vertical line marks the measured
    $\mathbf{Q}_\mathrm{SDW}=0.952\,H$~\cite{Fawcett1988}.
    (b) Fermi surface: (top) three-dimensional sheets coloured by the
    magnitude of the Fermi velocity $|\nabla_\mathbf{k}E|$, and (bottom)
    a band-resolved cross-section, showing the multi-armed electron jack
    centred at $\Gamma$ and the hole octahedra at $H$ whose parallel
    faces nest with the jack.
    (c) Interacting longitudinal spin susceptibility
    $\chi_{zz}(\mathbf{q})$ from the RPA Dyson equation for three
    interaction strengths $(U_{\uparrow\uparrow},U_{\uparrow\downarrow})$,
    in Ry.
    (d) Total density of states and the Cr~$4s$, $4p$ and $3d$ projections;
$\varepsilon_F$ lies in the pseudogap between bonding and antibonding $3d$
states, and $N(\varepsilon_F)$ is mainly of $3d$ character. All results are for a thermal smearing width $W = 0.005$~Ry ($T \approx 790$~K).}
  \label{fig:cr}
\end{figure*}

The density of states of nonmagnetic Cr [Fig.~\ref{fig:cr}(d)] is mainly
of Cr~$3d$ character, with $\varepsilon_F$ in the pseudogap between
the bonding and antibonding $d$ states. The Fermi surface
[Fig.~\ref{fig:cr}(b)] shows the jack at $\Gamma$ and the octahedra at $H$
with nearly parallel faces~\cite{Lomer1962}; both sheets are mainly of
$t_{2g}$ character~\cite{Heil2014}.

The bare susceptibility [Fig.~\ref{fig:cr}(a)] rises from about
$9~\mathrm{Ry}^{-1}$ at $\Gamma$ to a single maximum of about
$32~\mathrm{Ry}^{-1}$ at $H$, which a search over the whole zone confirms as
the global maximum. The interaction raises this peak [Fig.~\ref{fig:cr}(c)]:
$\chi_{zz}(H)$ grows from about $11~\mathrm{Ry}^{-1}$ at
$U_{\uparrow\uparrow}=0.020$~Ry to about $56~\mathrm{Ry}^{-1}$ at $0.060$~Ry.
With $\chi^0_\sigma(H)=\tfrac12\chi^0_{nn}(H)\approx16~\mathrm{Ry}^{-1}$, the
largest coupling, $I=0.054$~Ry, gives $I\chi^0_\sigma(H)\approx0.86$, and
$\chi_{zz}(H)$ would diverge at $I_c=1/\chi^0_\sigma(H)\approx0.063$~Ry. 
This
is close to the Stoner parameters of Cr from band calculations compiled in Table~X of Ref.~\onlinecite{Fawcett1988}, $0.051$--$0.068$~Ry, defined there, as here, by the exchange splitting per unit moment, which give
$I\chi^0_\sigma(H)=0.81$--$1.08$; the nonmagnetic state is therefore close
to an instability at $H$.
The uniform channel is far from this condition, with
$I_c=1/\chi^0_\sigma(\Gamma)\approx0.215$~Ry, so the instability is selective
in $\mathbf{q}$. Sandratskii and K\"ubler likewise found $\chi^0$ near $H$
about three times its value at $\mathbf{q}=0$~\cite{Sandratskii1992}; our
ratio is 3.4. Lowering the smearing to $W=0.001$~Ry ($158$~K) changes
$\chi^0_{nn}(H)$ by less than $2\%$ (Supplemental Material), so $I_c(H)$ is
insensitive to the electronic temperature in this range.

From Fig.~\ref{fig:cr}(c) it is clear that the calculated maximum lies at the commensurate point $H$, not at the experimental wave vector $\mathbf{Q}_\mathrm{SDW}$. Our
real-space sum reaches $r_\mathrm{cut}=33.0$~a.u., about six lattice
constants, which cannot resolve a modulation with a period of about 21
lattice constants; lowering the smearing does not move the maximum
(Supplemental Material).
Heil \emph{et al.} resolved the incommensurate peak in a non-spin-polarised
calculation with tetrahedron integration~\cite{Heil2014}, suggesting that $\delta$ is
already contained in the nonmagnetic bare susceptibility, as in Lomer's
argument~\cite{Lomer1962}. The here calculated response in the spin-polarised case is discussed in more detail below. 

\subsection{FeSe}
\label{subsec:fese}

FeSe is tetragonal ($P4/nmm$, Table~\ref{tab:struct}) and is the structurally
simplest iron-based superconductor~\cite{Hsu2008}. Below $T_s\approx90$~K it becomes orthorhombic through an electronic nematic
transition, which breaks the fourfold rotational symmetry of the Fe
plane~\cite{McQueen2009,WatsonPRB2015}; it superconducts below $T_c\approx8$~K, but
unlike most iron-based parent compounds it does not order magnetically at
ambient pressure~\cite{McQueen2009,Bohmer2018}. Its Fermi surface consists of
small hole pockets at $\Gamma$ and electron pockets at $M$ (two-Fe cell), of
mainly Fe~$3d_{xz/yz}$ and $3d_{xy}$ character~\cite{Heil2014,Maletz2014,WatsonPRB2015}.
Inelastic neutron scattering finds strong spin fluctuations at the stripe vector
$\mathbf{Q}_\mathrm{s}=(\pi,0)$ of the one-Fe cell, the ordering vector of the stripe
antiferromagnetic iron-based compounds, although FeSe itself does not
order~\cite{Wang2016}. Correlations are
strongly orbital dependent and shrink the pockets relative to
DFT results~\cite{Kreisel2020,Maletz2014}.

\begin{figure*}[htbp]
  \centering
  \includegraphics[width=\textwidth]{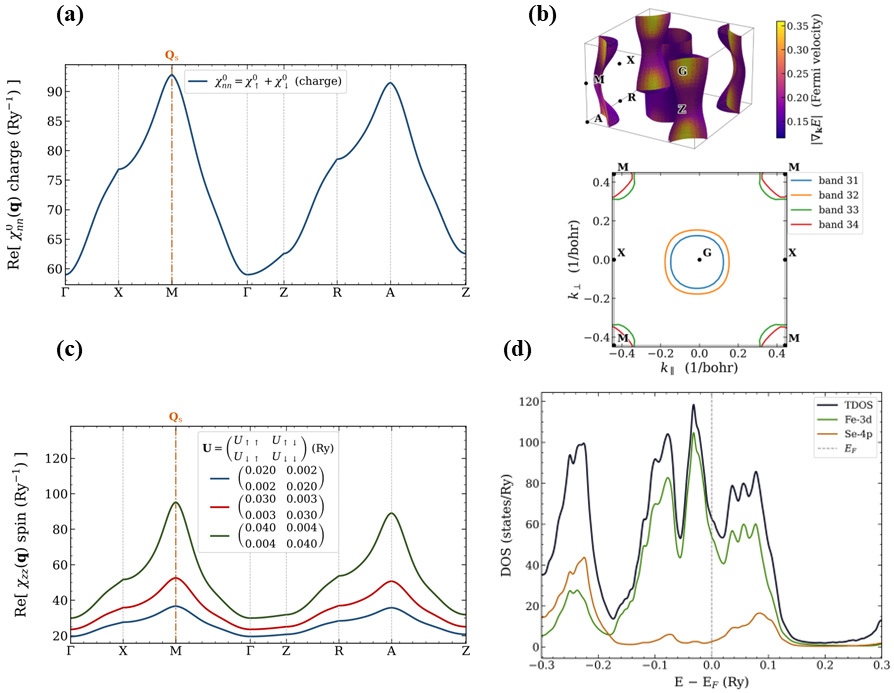}
  \caption{Results for FeSe.
    (a) Bare susceptibility $\chi^0_{nn}(\mathbf{q})$, whose $\mathbf{q}$-dependence
    coincides with that of the bare spin channel
    ($\chi^0_{zz}=\chi^0_{nn}/4$), along
    $\Gamma$-$X$-$M$-$\Gamma$-$Z$-$R$-$A$-$Z$, with $X=(\tfrac12,0,0)$,
    $M=(\tfrac12,\tfrac12,0)$, $Z=(0,0,\tfrac12)$, $R=(\tfrac12,0,\tfrac12)$
    and $A=(\tfrac12,\tfrac12,\tfrac12)$ in units of $(2\pi/a,2\pi/a,2\pi/c)$.
    The vertical line marks the stripe vector $\mathbf{Q}_\mathrm{s}$ at $M$, $(\pi,0)$ of the
    one-Fe lattice.
    (b) Calculated Fermi surface in the tetragonal (2-Fe) Brillouin
    zone: (top) three-dimensional sheets coloured by the magnitude of the
    Fermi velocity $|\nabla_\mathbf{k}E|$, and (bottom) a band-resolved
    cross-section, showing the concentric hole cylinders at $\Gamma$ and
    the electron cylinders at the zone corner $M$.
    (c) Interacting longitudinal spin susceptibility
    $\chi_{zz}(\mathbf{q})$ from the RPA Dyson equation for three
    interaction strengths $(U_{\uparrow\uparrow},U_{\uparrow\downarrow})$,
    in Ry.
    (d) Total density of states together with the Fe~$3d$ and Se~$4p$
    projections; the Fermi level sits on the upper shoulder of a large
    Fe~$3d$ manifold, giving a sizeable $N(\varepsilon_F)$ of
    predominantly Fe~$3d$ character. All results are for a thermal smearing width $W = 0.005$~Ry ($T \approx 790$~K).}
  \label{fig:fese}
\end{figure*}

Fe~$3d$ states dominate the density of states near $\varepsilon_F$
[Fig.~\ref{fig:fese}(d)]. The Fermi surface [Fig.~\ref{fig:fese}(b)] has hole
cylinders at $\Gamma$ and electron cylinders at $M$, as in ARPES, but with
larger pockets, as usual for DFT~\cite{Maletz2014}.

The bare susceptibility [Fig.~\ref{fig:fese}(a)] has maxima of about
$93~\mathrm{Ry}^{-1}$ at $M$ and $91~\mathrm{Ry}^{-1}$ at $A$, and a minimum of
about $59~\mathrm{Ry}^{-1}$ at $\Gamma$; the leading eigenvalue of the site
matrix peaks at $M$. Without Se~$4p$, that is with the Fe--Fe terms alone, $\chi^0_{nn}$ drops by about $8\%$ at $M$, to $86~\mathrm{Ry}^{-1}$, hardly changes at $\Gamma$ ($60~\mathrm{Ry}^{-1}$), and keeps its maximum at $M$; $\lambda_\mathrm{max}$ involves only the Fe sites and does not change.
$M$ connects the hole pockets at $\Gamma$ with the electron pockets at $M$ and
corresponds to the stripe vector $(\pi,0)$ of the one-Fe cell; $A$ is the same
in-plane vector with $q_z=\pi/c$. Heil \emph{et al.} also obtained the maximum
at $M$, but only when bands away from $\varepsilon_F$ were
included~\cite{Heil2014}.

The interaction raises the peaks at $M$ and $A$ [Fig.~\ref{fig:fese}(c)]:
$\chi_{zz}(M)$ grows from about $37~\mathrm{Ry}^{-1}$ at
$U_{\uparrow\uparrow}=0.020$~Ry per Fe to about $95~\mathrm{Ry}^{-1}$ at
$0.040$~Ry, where $I\lambda_\mathrm{max}(M)\approx0.77$. 
As Figs.~\ref{fig:fese}(a) and~\ref{fig:fese}(c) show, the calculated maximum lies at $M$, the wave vector of the observed stripe spin fluctuations. 
The couplings listed above are clearly
below the Stoner parameter of Fe: with $I\approx0.068$~Ry per Fe
atom~\cite{Gunnarsson1976,Janak1977}, $I\lambda_\mathrm{max}(M)\approx1.5$, and the
nonmagnetic state would be unstable at $M$. This is the known tendency of DFT
to stabilise stripe order in FeSe, which is not
observed~\cite{Bohmer2018,Wang2016}. The calculation identifies the wave
vector of the dominant spin fluctuations, but not whether they order.
In both FeSe and bcc~Cr the ordering vector is already the maximum of the bare susceptibility. Among the CDW systems the maximum lies close to the observed vector in LaTe$_3$, at its image $\mathbf{Q}'_\mathrm{CDW}$, but not in 2H-NbSe$_2$, where the bare response does not by itself fix the ordering vector.

\subsection{Longitudinal response in the magnetically ordered state}
\label{sec:ordered}

\begin{figure*}[htbp]
  \centering
 \includegraphics[width=\textwidth]{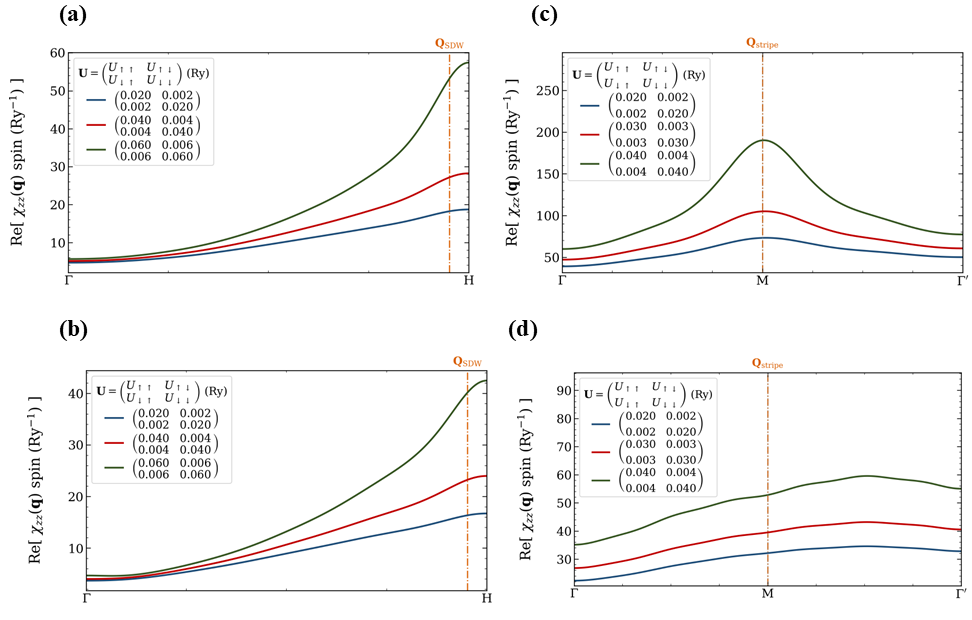}
\caption{Longitudinal spin susceptibility $\chi_{zz}(\mathbf{q})$ from the
RPA Dyson equation in the nonmagnetic and magnetically ordered states.
(a) bcc~Cr in the nonmagnetic two-atom cell and (b) in the G-type
antiferromagnetic state of the same cell, along $\Gamma$-$H$, for
$U_{\uparrow\uparrow}=0.020$, $0.040$ and $0.060$~Ry per Cr.
(c) FeSe in the nonmagnetic $\sqrt{2}a\times\sqrt{2}a\times c$ supercell and
(d) in the single-stripe antiferromagnetic state of the same cell, along
$\Gamma$-$M$-$\Gamma'$, where $M=(\tfrac12,-\tfrac12,0)$ in units of $2\pi/a$
of the two-Fe cell is the stripe ordering vector $\mathbf{Q}_\mathrm{s}$ and $\Gamma'=2M$ (equivalent by symmetry to $M$ of Fig.~\ref{fig:fese}), for $U_{\uparrow\uparrow}=0.020$, $0.030$ and $0.040$~Ry per Fe.
Throughout $U_{\downarrow\downarrow}=U_{\uparrow\uparrow}$ and
$U_{\uparrow\downarrow}=0.1\,U_{\uparrow\uparrow}$ (see text).
Susceptibilities are per unit cell of the magnetic cell. Magnetic order
reduces $\chi_{zz}$ at all wave vectors in both systems; at the ordering
vector the bare response drops by about 8\% in Cr and by about half in FeSe, and the interacting response by
more. In Cr the maximum stays at $H$; in FeSe it moves from
$M$ to about $1.5M$, the boundary of the magnetic Brillouin zone.
All results are for a thermal smearing width $W = 0.005$~Ry ($T \approx 790$~K).
}
\label{fig:afm}
\end{figure*}

All previous results were calculated for the nonmagnetic state, where
$\chi^0_\uparrow = \chi^0_\downarrow$ and Eq.~\eqref{eq:chizz} reduce to a
single Stoner-like channel. This means we were only testing how close the
nonmagnetic state is to becoming unstable. We now ask what happens to the
longitudinal response once the material actually orders magnetically.
For bcc~Cr we model the spin-density wave as a simple G-type (CsCl-like)
antiferromagnet, with Cr$_\uparrow$ at $(0,0,0)$ and Cr$_\downarrow$ at
$(\tfrac{1}{2},\tfrac{1}{2},\tfrac{1}{2})$ in the conventional two-atom cell. For FeSe we impose single-stripe order at $\mathbf{Q}_\mathrm{s}=(\pi,0)$ in a
$\sqrt{2}a\times\sqrt{2}a\times c$ supercell with four Fe and four Se atoms, where each Fe has two nearest-neighbour Fe of each spin and the total cell magnetic moment is zero. For FeSe the structural parameters are unchanged; for bcc~Cr both panels use
the cell described in Sec.~\ref{subsec:compdet}. We keep the labels of the
crystallographic cells, so that $H$ and $M$ denote the same or symmetry-equivalent wave vectors as in
Figs.~\ref{fig:cr} and~\ref{fig:fese}. In the magnetic cells both fold onto
$\Gamma$, and the four FeSe points $\Gamma$, $M=(\pi,0)$, $(0,\pi)$ and
$(\pi,\pi)$ become equivalent in the reduced zone. With the bond phase of
Eq.~\eqref{eq:ft} they remain distinct.

Since $\chi^0$ is defined per unit cell [Eq.~\eqref{eq:sumrule}], a cell with
$N$ formula units has $N$ times the bubble of the primitive cell. The kernel
acts on each site, so the same $U_{\sigma\sigma'}$ applies in both cells and
the results do not depend on the choice of cell: for bcc~Cr at the
experimental volume, $I=0.054$~Ry per Cr gives $I\lambda_\mathrm{max}(H)=0.86$
in both the one- and the two-atom cell. For FeSe we use
$U_{\uparrow\uparrow}=0.020$, $0.030$ and $0.040$~Ry per Fe, as in
Sec.~\ref{subsec:fese}, which give $I\lambda_\mathrm{max}(M)=0.39$, $0.58$ and
$0.77$. For Cr the two-atom cell is at a smaller volume, and
$U_{\uparrow\uparrow}=0.020$, $0.040$ and $0.060$~Ry per Cr give
$I\lambda_\mathrm{max}(H)=0.25$, $0.50$ and $0.76$ in its nonmagnetic state. We use the same interaction kernel for both nonmagnetic and magnetic calculations of each system. Absolute $\chi^0$ values cannot be compared between cells of different size because doubling the formula units doubles $N(\varepsilon_F)$; after accounting for this, the nonmagnetic supercell results agree with the earlier ones to $0.2\%$; for bcc~Cr the two-atom cell is at a different volume (Sec.~\ref{subsec:compdet}), so its nonmagnetic panel serves as the reference for the ordered one. Within the same cell, the two panels of Fig.~\ref{fig:afm} can be compared directly.
Here $\chi^0_{zz}=\tfrac14\chi^0_{nn}=\tfrac12\chi^0_\sigma$ per magnetic
cell; for FeSe this is twice the value in the two-Fe cell of
Sec.~\ref{subsec:fese} ($23.2~\mathrm{Ry}^{-1}$ at $M$).

For bcc~Cr [Figs.~\ref{fig:afm}(a) and~\ref{fig:afm}(b)] the peak remains at
$H$, but $\chi^0_{zz}(H)$ drops from $14.0$ to $12.8~\mathrm{Ry}^{-1}$. The
suppression depends on momentum: at $\Gamma$, $\chi^0_{zz}$ drops from $4.3$
to $3.4~\mathrm{Ry}^{-1}$, so the ratio $\chi^0_{zz}(H)/\chi^0_{zz}(\Gamma)$
increases from $3.2$ to $3.7$. The staggered response at $H$, which
corresponds to the order parameter, thus remains the main feature, though
reduced.

FeSe behaves differently [Figs.~\ref{fig:afm}(c) and~\ref{fig:afm}(d)]. In
the nonmagnetic supercell the response along $\Gamma$-$M$-$\Gamma'$ is twice
that of the two-Fe cell, with $\chi^0_{zz}(M)=46.4$ and
$\chi^0_{zz}(\Gamma)=29.5~\mathrm{Ry}^{-1}$ (ratio $1.57$). In the stripe state
the response drops by $40$ to $50\%$ along the whole line, to $24.2$ at $M$
and $17.6~\mathrm{Ry}^{-1}$ at $\Gamma$, and its maximum moves from $M$ to about
$1.5M$ ($25.6~\mathrm{Ry}^{-1}$). This is the boundary of the magnetic Brillouin
zone along the stripe direction.

Both reductions are consistent with exchange splitting removing states from
the Fermi level: $\chi^0_{zz}(\Gamma)$, a measure of the thermally averaged
density of states [Eq.~\eqref{eq:sumrule}], drops by about $20\%$ in Cr and
$40\%$ in FeSe. In Cr the order is expected to gap the nested sheets that
produce the maximum at $H$~\cite{Lomer1962,Heil2014}. In FeSe the reduction is
broad, and the maximum moves away from the ordering vector to the boundary of
the magnetic zone. Some suppression is always
expected: in the ordered phase the longitudinal fluctuation at $\mathbf{Q}$
becomes the massive amplitude mode of the order parameter, while the divergence
in Eq.~\eqref{eq:dyson} describes approaching the transition from the
nonmagnetic side.

In the ordered states the two sublattices have different spin bubbles, and
the site-resolved Dyson equation (Sec.~\ref{subsec:dyson}) treats them
explicitly. In both ordered states we take
$U_{\downarrow\downarrow}=U_{\uparrow\uparrow}$; since the diagonal
elements differ on each site in a spin-polarised system
(Sec.~\ref{subsec:dyson}), this is an approximation. Finally, density functional theory overestimates
the stripe moment in FeSe, which has no long-range order at ambient pressure
\cite{Bohmer2018,Wang2016}. The stripe state we impose is therefore a reference
configuration, not the true ground state, and the suppression we report is probably overestimated.

\section{Orbital character of the bare susceptibility}
\label{sec:orbital}

\begin{table}[t]
\caption{Orbitals that carry the bare susceptibility. $R$ is the length of
the bond with the largest element sum $|\sum_{LL'}\chi^0_{LL'}(\mathbf{R})|$;
the last column lists the orbitals with the largest weight in the leading
eigenvector of the symmetrised matrix on that bond, and for NbSe$_2$ both
members of the degenerate pair $d_{x^2-y^2}$, $d_{xy}$. The orbitals refer to
the crystal axes, except for FeSe, where they refer to the local axes of the
Fe site (primed), with $z'$ along an in-plane nearest-neighbour Fe--Fe bond. The full matrices are given
in the Supplemental Material.}
\label{tab:orbital_summary}
\begin{ruledtabular}
\begin{tabular}{lccl}
System & Orbitals & $R$ (a.u.) & Leading orbitals \\
\hline
2H-NbSe$_2$ & Nb $4d$ & 6.50 & $d_{z^2}$, $d_{x^2-y^2}$, $d_{xy}$ (mixed) \\
LaTe$_3$    & Te $5p$ & 5.89 & $p_x$, $p_z$ \\
bcc Cr      & Cr $3d$ & 4.72 & $t_{2g}$, mixed with $e_g$ \\
FeSe        & Fe $3d$ & 7.12 & $d_{x'y'}$, $d_{x'^2-y'^2}$ \\
\end{tabular}
\end{ruledtabular}
\end{table}

The results above show where in reciprocal space each material responds, but
not which orbitals carry the response. To see this we keep the orbital
indices in Eq.~\eqref{eq:chi0} and examine the matrix
$\chi^0_{\sigma,LL'}(\mathbf{R})$ on individual bonds. The sum of all its
elements is the bond contribution to Eq.~\eqref{eq:chi0}. Diagonal elements are
contributions of single orbitals, and off-diagonal elements are contributions
of pairs of different orbitals. Only the sum is independent of the choice of
local axes; we quote individual elements in the crystal frame, except for FeSe (Table~\ref{tab:orbital_summary}). Since
$\chi^0_{LL'}(\mathbf{R})=\chi^0_{L'L}(-\mathbf{R})$, the matrix on a bond need
not be symmetric. The decomposition requires a Green's-function cluster with a
single $\ell$, so we use Nb~$4d$, Te~$5p$, Cr~$3d$ (two-atom cubic cell) and
Fe~$3d$ ($\sqrt2a\times\sqrt2a\times c$ cell) clusters. The results are
summarised in Table~\ref{tab:orbital_summary}.

In bcc~Cr the response on the nearest-neighbour $\langle111\rangle$ bond comes
mainly from the $t_{2g}$ orbitals, and the elements that mix $t_{2g}$ with
$e_g$ are as large as the diagonal ones. On the second-neighbour
$\langle100\rangle$ bond the $t_{2g}$--$e_g$ elements vanish, as the fourfold
symmetry of this bond requires (Supplemental Material), and the response has
the opposite sign.
At $H$ the Fourier phase is $-1$ for $\langle111\rangle$ bonds and $+1$ for
$\langle100\rangle$ bonds, so the two shells add. Of these two shells only the
$\langle111\rangle$ shell distinguishes $H$ from $\Gamma$, since the
$\langle100\rangle$ phase is the same at both; about half of it comes from the
$t_{2g}$ block and most of the rest from $t_{2g}$--$e_g$ mixing (Supplemental
Material), consistent with the $t_{2g}$ character of the nesting
sheets~\cite{Heil2014}.

In 2H-NbSe$_2$ the diagonal elements on the in-plane nearest-neighbour bond
nearly cancel, and the response comes from mixing between $d_{z^2}$,
$d_{x^2-y^2}$ and $d_{xy}$. The $d_{xz}$ and $d_{yz}$ orbitals, which are odd
under the mirror plane containing the Nb layer, do not mix with these three on
any in-plane bond, as the mirror symmetry requires. On the bond between layers
along $c$ the response is carried almost entirely by $d_{z^2}$, as expected
for $\sigma$ bonding along $c$.

In LaTe$_3$ the response comes almost entirely from the two Te sites of the
nets; the Te site of the LaTe layer contributes little, both on site and on
its bonds. On the nearest-neighbour Te--Te bond of the net, which runs along
$(\mathbf{a}+\mathbf{c})/2$, $p_x$ and $p_z$ contribute equally and the
off-diagonal elements are as large as the diagonal ones, so the response is
carried not by $p_x$ or $p_z$ alone but by their combinations along the Te--Te
bonds, the chain orbitals of the tight-binding description of
Ref.~\onlinecite{Brouet2008}, whose $p_x$ and $p_z$ axes lie along the Te--Te
bonds. The out-of-plane $p_y$ orbital contributes little and hardly
mixes with $p_x$ and $p_z$. These are the in-plane states that form the flat
sheets of Fig.~\ref{fig:late3}(b)~\cite{Brouet2004,Gweon1998}.

In FeSe the in-plane bonds involve several $d$ orbitals with comparable
weight, and no single orbital dominates. The response on the Fe--Fe bond
along $c$ is more than an order of magnitude weaker than on the in-plane
bonds, so it is essentially two-dimensional, consistent with the stripe
fluctuations at $\mathbf{Q}_\mathrm{s}=(\pi,0)$~\cite{Wang2016}.

In all four materials the orbital character found in this way agrees with the
projected density of states and with the orbital character of the Fermi
surfaces reported in Refs.~\onlinecite{Heil2014,Johannes2006,Gweon1998}. What the
decomposition adds is the size of the interorbital terms. On the
nearest-neighbour bonds of Cr, NbSe$_2$ and LaTe$_3$ they are as large as, or
larger than, the diagonal terms, whereas on bonds where symmetry forbids
mixing, such as $\langle100\rangle$ in Cr, they vanish. A treatment that keeps
only the diagonal, orbital-by-orbital contributions to $\chi^0$ misses this.

\section{Summary}
\label{sec:summary}

We have developed a real-space formulation of the bare longitudinal susceptibility, and from it the interacting one, in a full-potential, all-electron electronic structure method, following the formalism of Katsnelson and
Lichtenstein~\cite{katsnelson2004magnetic}. The bare bubbles are built in real space from Matsubara Green's functions and resolved by site, spin and orbital degrees of freedom. Further, the Dyson equation with an on-site exchange-correlation kernel is solved in
reciprocal space at each $\mathbf{q}$.

As a test, we demonstrate that the method reproduces the well-known temperature-dependent Stoner
criterion for bcc~Fe, in agreement with the results of Gunnarsson~\cite{Gunnarsson1976}.
For bcc~Cr the bare susceptibility has its maximum at the
commensurate ordering vector $H$, while for FeSe the interacting response peaks at $M$, the wave vector of the observed stripe spin fluctuations.
For Cr the critical Stoner parameter at $H$ is close to the Stoner parameters of Cr from band calculations~\cite{Fawcett1988}.
In LaTe$_3$ the response of the Te nets peaks close to the observed charge-density-wave vector, at its image $\tfrac57\,(2\pi/c)$ in the neighbouring Brillouin zone, the vector that nests the Fermi surface of a single Te net. Which of the two equivalent vectors carries the response is set by the phases of the sites in the cell, as for neutron scattering intensities in different Brillouin zones~\cite{Lovesey1984}. In 2H-NbSe$_2$ the bare response shows several maxima of similar height, and $\mathbf{Q}_\mathrm{CDW}$ lies on the flank of the largest, in agreement with earlier work that assigns its selection to electron-phonon coupling~\cite{Johannes2006,Weber2011}.
Antiferromagnetic order reduces the bare longitudinal response at the ordering vector, by about 8\% in Cr and by about half in FeSe; in Cr the maximum stays at $H$, whereas in FeSe it moves to
the boundary of the magnetic Brillouin zone. The orbital decomposition shows
that interorbital terms are as large as the diagonal ones on the leading
bonds of Cr, NbSe$_2$ and LaTe$_3$ in the crystal frame.

\section*{Data Availability}
The data supporting the findings of this study, including RSPt input files
and tabulated susceptibility data for all four benchmark systems, are
available from the corresponding authors upon reasonable request.

\section*{Code Availability}
The modified RSPt routines and the post-processing code used to compute the
susceptibilities are available from the corresponding authors upon
reasonable request. RSPt is freely available at
\url{https://github.com/uumaterialstheory/rspt}.

\section*{Use of AI}
Claude (Anthropic) was used to help write the post-processing and plotting scripts, modify the RSPt code to calculate the $U$ matrix [Eq.~\eqref{eq:Ukernel}], and edit the text; the authors checked all output and take full responsibility for the content.

\section*{Acknowledgements}

V.B.\ acknowledges financial support from the Swedish Research Council
(Vetenskapsr{\aa}det, VR; Grant No.\ 2024-05206), and A.D.\ from the Swedish
Research Council (VR; Grant No.\ 2024-04986). O.E.\ and A.D.\ acknowledge
support from the Knut and Alice Wallenberg Foundation (KAW; Grant
No.\ 2022.0108) and from the Wallenberg Initiative Materials Science for
Sustainability (WISE), funded by the Knut and Alice Wallenberg Foundation.
A.D.\ also acknowledges the Swedish e-Science Research Centre (SeRC).
R.P.\ and V.B.\ acknowledge support from the G\"oran Gustafsson Foundation
(recipient of the ``small prize'': V.B.).
O.E.\ also acknowledges support from the Swedish Research Council (VR), the
Foundation for Strategic Research (SSF), the European Research Council (854843-FASTCORR),
Knut and Alice Wallenberg Foundation (KAW scholar), 
eSSENCE and STandUP. M.I.K.\ acknowledges support from the Wallenberg Initiative Materials Science for Sustainability (WISE) funded by the Knut and Alice Wallenberg Foundation (KAW) and the European Research Council through the ERC Synergy Grant 854843-FASTCORR.

Computational resources were provided by the National Academic
Infrastructure for Supercomputing in Sweden (NAISS) at the National
Supercomputer Centre (NSC, Tetralith) and the PDC Center for High Performance
Computing (Dardel), funded by the Swedish Research Council.

\section*{Author Contributions}
O.E.\ and M.I.K.\ conceived the project. R.P.\ implemented the modified RSPt routines, performed all calculations
and wrote the original version of the manuscript. A.M.\ wrote the Python implementation of the RPA
Dyson solver. V.B.\ and O.E.\ wrote the routines that calculate the $U$ matrices used in this work. V.B., M.I.K., A.D.\ and O.E.\ supervised the project and suggested improvements. All authors discussed
the results and commented on the manuscript.

\section*{Competing Interests}
The authors declare no competing interests.

\medskip
\bibliographystyle{naturemag.bst}
\bibliography{main}

\begin{thebibliography}{10}
\expandafter\ifx\csname url\endcsname\relax
  \def\url#1{\texttt{#1}}\fi
\expandafter\ifx\csname urlprefix\endcsname\relax\def\urlprefix{URL }\fi
\providecommand{\bibinfo}[2]{#2}
\providecommand{\eprint}[2][]{\url{#2}}

\bibitem{Gruner1988}
\bibinfo{author}{Gr{\"u}ner, G.}
\newblock \emph{\bibinfo{title}{Density Waves in Solids}}
  (\bibinfo{publisher}{Addison-Wesley}, \bibinfo{address}{Reading, MA},
  \bibinfo{year}{1994}).

\bibitem{Overhauser1962}
\bibinfo{author}{Overhauser, A.~W.}
\newblock \bibinfo{title}{Spin density waves in an electron gas}.
\newblock \emph{\bibinfo{journal}{Physical Review}}
  \textbf{\bibinfo{volume}{128}}, \bibinfo{pages}{1437--1452}
  (\bibinfo{year}{1962}).

\bibitem{Zhu2015}
\bibinfo{author}{Zhu, X.}, \bibinfo{author}{Cao, Y.}, \bibinfo{author}{Zhang,
  J.}, \bibinfo{author}{Plummer, E.~W.} \& \bibinfo{author}{Guo, J.}
\newblock \bibinfo{title}{Classification of charge density waves based on their
  nature}.
\newblock \emph{\bibinfo{journal}{Proc. Natl. Acad. Sci. USA}}
  \textbf{\bibinfo{volume}{112}}, \bibinfo{pages}{2367} (\bibinfo{year}{2015}).

\bibitem{Keimer2015}
\bibinfo{author}{Keimer, B.}, \bibinfo{author}{Kivelson, S.~A.},
  \bibinfo{author}{Norman, M.~R.}, \bibinfo{author}{Uchida, S.} \&
  \bibinfo{author}{Zaanen, J.}
\newblock \bibinfo{title}{From quantum matter to high-temperature
  superconductivity in copper oxides}.
\newblock \emph{\bibinfo{journal}{Nature}} \textbf{\bibinfo{volume}{518}},
  \bibinfo{pages}{179} (\bibinfo{year}{2015}).

\bibitem{Fernandes2022}
\bibinfo{author}{Fernandes, R.~M.} \emph{et~al.}
\newblock \bibinfo{title}{Iron pnictides and chalcogenides: a new paradigm for
  superconductivity}.
\newblock \emph{\bibinfo{journal}{Nature}} \textbf{\bibinfo{volume}{601}},
  \bibinfo{pages}{35} (\bibinfo{year}{2022}).

\bibitem{Rossnagel2011}
\bibinfo{author}{Rossnagel, K.}
\newblock \bibinfo{title}{On the origin of charge-density waves in select
  layered transition-metal dichalcogenides}.
\newblock \emph{\bibinfo{journal}{Journal of Physics: Condensed Matter}}
  \textbf{\bibinfo{volume}{23}}, \bibinfo{pages}{213001}
  (\bibinfo{year}{2011}).

\bibitem{Fernandes2014}
\bibinfo{author}{Fernandes, R.~M.}, \bibinfo{author}{Chubukov, A.~V.} \&
  \bibinfo{author}{Schmalian, J.}
\newblock \bibinfo{title}{What drives nematic order in iron-based
  superconductors?}
\newblock \emph{\bibinfo{journal}{Nat. Phys.}} \textbf{\bibinfo{volume}{10}},
  \bibinfo{pages}{97} (\bibinfo{year}{2014}).

\bibitem{deMedici2014}
\bibinfo{author}{de' Medici, L.}, \bibinfo{author}{Giovannetti, G.} \&
  \bibinfo{author}{Capone, M.}
\newblock \bibinfo{title}{Selective {Mott} physics as a key to iron
  superconductors}.
\newblock \emph{\bibinfo{journal}{Phys. Rev. Lett.}}
  \textbf{\bibinfo{volume}{112}}, \bibinfo{pages}{177001}
  (\bibinfo{year}{2014}).

\bibitem{Yi2019}
\bibinfo{author}{Yi, M.} \emph{et~al.}
\newblock \bibinfo{title}{Nematic energy scale and the missing electron pocket
  in {FeSe}}.
\newblock \emph{\bibinfo{journal}{Phys. Rev. X}} \textbf{\bibinfo{volume}{9}},
  \bibinfo{pages}{041049} (\bibinfo{year}{2019}).

\bibitem{Johnston2010}
\bibinfo{author}{Johnston, D.~C.}
\newblock \bibinfo{title}{The puzzle of high temperature superconductivity in
  layered iron pnictides and chalcogenides}.
\newblock \emph{\bibinfo{journal}{Advances in Physics}}
  \textbf{\bibinfo{volume}{59}}, \bibinfo{pages}{803--1061}
  (\bibinfo{year}{2010}).

\bibitem{Bohmer2018}
\bibinfo{author}{B{\"o}hmer, A.~E.} \& \bibinfo{author}{Kreisel, A.}
\newblock \bibinfo{title}{Nematicity, magnetism and superconductivity in
  {FeSe}}.
\newblock \emph{\bibinfo{journal}{Journal of Physics: Condensed Matter}}
  \textbf{\bibinfo{volume}{30}}, \bibinfo{pages}{023001}
  (\bibinfo{year}{2018}).

\bibitem{Peierls1955}
\bibinfo{author}{Peierls, R.~E.}
\newblock \emph{\bibinfo{title}{Quantum Theory of Solids}}
  (\bibinfo{publisher}{Oxford University Press}, \bibinfo{address}{Oxford},
  \bibinfo{year}{1955}).

\bibitem{Johannes2006}
\bibinfo{author}{Johannes, M.~D.}, \bibinfo{author}{Mazin, I.~I.} \&
  \bibinfo{author}{Howells, C.~A.}
\newblock \bibinfo{title}{Fermi-surface nesting and the origin of the
  charge-density wave in {NbSe$_2$}}.
\newblock \emph{\bibinfo{journal}{Physical Review B}}
  \textbf{\bibinfo{volume}{73}}, \bibinfo{pages}{205102}
  (\bibinfo{year}{2006}).

\bibitem{Irkhin2001}
\bibinfo{author}{Irkhin, V.~Y.}, \bibinfo{author}{Katanin, A.~A.} \&
  \bibinfo{author}{Katsnelson, M.~I.}
\newblock \bibinfo{title}{Effects of van {Hove} singularities on magnetism and
  superconductivity in the ${t\ensuremath{-}t}^{\ensuremath{'}}$ {Hubbard}
  model: A parquet approach}.
\newblock \emph{\bibinfo{journal}{Phys. Rev. B}} \textbf{\bibinfo{volume}{64}},
  \bibinfo{pages}{165107} (\bibinfo{year}{2001}).

\bibitem{Weber2011}
\bibinfo{author}{Weber, F.} \emph{et~al.}
\newblock \bibinfo{title}{Extended phonon collapse and the origin of the
  charge-density wave in {2H-NbSe$_2$}}.
\newblock \emph{\bibinfo{journal}{Phys. Rev. Lett.}}
  \textbf{\bibinfo{volume}{107}}, \bibinfo{pages}{107403}
  (\bibinfo{year}{2011}).

\bibitem{BohmPines1953}
\bibinfo{author}{Bohm, D.} \& \bibinfo{author}{Pines, D.}
\newblock \bibinfo{title}{A collective description of electron interactions:
  {III}. {Coulomb} interactions in a degenerate electron gas}.
\newblock \emph{\bibinfo{journal}{Phys. Rev.}} \textbf{\bibinfo{volume}{92}},
  \bibinfo{pages}{609--625} (\bibinfo{year}{1953}).

\bibitem{Fawcett1988}
\bibinfo{author}{Fawcett, E.}
\newblock \bibinfo{title}{Spin-density-wave antiferromagnetism in chromium}.
\newblock \emph{\bibinfo{journal}{Reviews of Modern Physics}}
  \textbf{\bibinfo{volume}{60}}, \bibinfo{pages}{209--283}
  (\bibinfo{year}{1988}).

\bibitem{Lomer1962}
\bibinfo{author}{Lomer, W.~M.}
\newblock \bibinfo{title}{Electronic structure of chromium group metals}.
\newblock \emph{\bibinfo{journal}{Proceedings of the Physical Society}}
  \textbf{\bibinfo{volume}{80}}, \bibinfo{pages}{489--496}
  (\bibinfo{year}{1962}).

\bibitem{Wang2016}
\bibinfo{author}{Wang, Q.} \emph{et~al.}
\newblock \bibinfo{title}{Strong interplay between stripe spin fluctuations,
  nematicity and superconductivity in {FeSe}}.
\newblock \emph{\bibinfo{journal}{Nature Materials}}
  \textbf{\bibinfo{volume}{15}}, \bibinfo{pages}{159--163}
  (\bibinfo{year}{2016}).

\bibitem{Kreisel2020}
\bibinfo{author}{Kreisel, A.}, \bibinfo{author}{Hirschfeld, P.~J.} \&
  \bibinfo{author}{Andersen, B.~M.}
\newblock \bibinfo{title}{On the remarkable superconductivity of {FeSe} and its
  close cousins}.
\newblock \emph{\bibinfo{journal}{Symmetry}} \textbf{\bibinfo{volume}{12}},
  \bibinfo{pages}{1402} (\bibinfo{year}{2020}).

\bibitem{moriya_book}
\bibinfo{author}{Moriya, T.}
\newblock \emph{\bibinfo{title}{{Spin Fluctuations in Itinerant Electron
  Magnetism}}} (\bibinfo{publisher}{Springer Science \& Business Media},
  \bibinfo{year}{2012}).

\bibitem{Wilson1975}
\bibinfo{author}{Wilson, J.~A.}, \bibinfo{author}{Di~Salvo, F.~J.} \&
  \bibinfo{author}{Mahajan, S.}
\newblock \bibinfo{title}{Charge-density waves and superlattices in the
  metallic layered transition metal dichalcogenides}.
\newblock \emph{\bibinfo{journal}{Advances in Physics}}
  \textbf{\bibinfo{volume}{24}}, \bibinfo{pages}{117--201}
  (\bibinfo{year}{1975}).

\bibitem{DiSalvo1976}
\bibinfo{author}{Di~Salvo, F.~J.}, \bibinfo{author}{Moncton, D.~E.} \&
  \bibinfo{author}{Waszczak, J.~V.}
\newblock \bibinfo{title}{Electronic properties and superlattice formation in
  the semimetal {TiSe$_2$}}.
\newblock \emph{\bibinfo{journal}{Physical Review B}}
  \textbf{\bibinfo{volume}{14}}, \bibinfo{pages}{4321--4328}
  (\bibinfo{year}{1976}).

\bibitem{Gweon1998}
\bibinfo{author}{Gweon, G.-H.} \emph{et~al.}
\newblock \bibinfo{title}{Direct observation of complete {Fermi} surface,
  imperfect nesting, and gap anisotropy in the high-temperature incommensurate
  charge-density-wave compound {SmTe$_3$}}.
\newblock \emph{\bibinfo{journal}{Physical Review Letters}}
  \textbf{\bibinfo{volume}{81}}, \bibinfo{pages}{886--889}
  (\bibinfo{year}{1998}).

\bibitem{Brouet2008}
\bibinfo{author}{Brouet, V.} \emph{et~al.}
\newblock \bibinfo{title}{Angle-resolved photoemission study of the evolution
  of band structure and charge density wave properties in {$R$Te$_3$} ({$R$} =
  {Y}, {La}, {Ce}, {Sm}, {Gd}, {Tb}, and {Dy})}.
\newblock \emph{\bibinfo{journal}{Physical Review B}}
  \textbf{\bibinfo{volume}{77}}, \bibinfo{pages}{235104}
  (\bibinfo{year}{2008}).

\bibitem{Ru2008}
\bibinfo{author}{Ru, N.} \emph{et~al.}
\newblock \bibinfo{title}{Effect of chemical pressure on the charge density
  wave transition in rare-earth tritellurides {$R$Te$_3$}}.
\newblock \emph{\bibinfo{journal}{Physical Review B}}
  \textbf{\bibinfo{volume}{77}}, \bibinfo{pages}{035114}
  (\bibinfo{year}{2008}).

\bibitem{Hong2022}
\bibinfo{author}{Hong, Y.}, \bibinfo{author}{Wei, Q.}, \bibinfo{author}{Liang,
  X.} \& \bibinfo{author}{Lu, W.}
\newblock \bibinfo{title}{Origin and strain tuning of charge density wave in
  {LaTe$_3$}}.
\newblock \emph{\bibinfo{journal}{Physica B}} \textbf{\bibinfo{volume}{639}},
  \bibinfo{pages}{413988} (\bibinfo{year}{2022}).

\bibitem{Samolyuk2009}
\bibinfo{author}{Samolyuk, G.~D.} \& \bibinfo{author}{Antropov, V.~P.}
\newblock \bibinfo{title}{Character of magnetic instabilities in
  {CaFe$_2$As$_2$}}.
\newblock \emph{\bibinfo{journal}{Phys. Rev. B}} \textbf{\bibinfo{volume}{79}},
  \bibinfo{pages}{052505} (\bibinfo{year}{2009}).

\bibitem{Heil2014}
\bibinfo{author}{Heil, C.}, \bibinfo{author}{Sormann, H.},
  \bibinfo{author}{Boeri, L.}, \bibinfo{author}{Aichhorn, M.} \&
  \bibinfo{author}{von~der Linden, W.}
\newblock \bibinfo{title}{Accurate bare susceptibilities from full-potential ab
  initio calculations}.
\newblock \emph{\bibinfo{journal}{Physical Review B}}
  \textbf{\bibinfo{volume}{90}}, \bibinfo{pages}{115143}
  (\bibinfo{year}{2014}).

\bibitem{JohannesMazin2008}
\bibinfo{author}{Johannes, M.~D.} \& \bibinfo{author}{Mazin, I.~I.}
\newblock \bibinfo{title}{Fermi surface nesting and the origin of charge
  density waves in metals}.
\newblock \emph{\bibinfo{journal}{Physical Review B}}
  \textbf{\bibinfo{volume}{77}}, \bibinfo{pages}{165135}
  (\bibinfo{year}{2008}).

\bibitem{Ekuma2025NESTOR}
\bibinfo{author}{Nwaogbo, C.~I.} \& \bibinfo{author}{Ekuma, C.~E.}
\newblock \bibinfo{title}{{NESTOR}: An open-source computational toolkit for
  electronic instabilities}.
\newblock \bibinfo{howpublished}{Preprint, SSRN} (\bibinfo{year}{2025}).
\newblock \urlprefix\url{https://dx.doi.org/10.2139/ssrn.5718304}.

\bibitem{Wills2010}
\bibinfo{author}{Wills, J.~M.} \emph{et~al.}
\newblock \bibinfo{title}{Full-potential electronic structure method: Energy
  and force calculations with density functional and dynamical mean field
  theory}.
\newblock \emph{\bibinfo{journal}{Springer Series in Solid-State Sciences}}
  \textbf{\bibinfo{volume}{167}} (\bibinfo{year}{2010}).

\bibitem{katsnelson2004magnetic}
\bibinfo{author}{Katsnelson, M.~I.} \& \bibinfo{author}{Lichtenstein, A.~I.}
\newblock \bibinfo{title}{Magnetic susceptibility, exchange interactions and
  spin-wave spectra in the local spin density approximation}.
\newblock \emph{\bibinfo{journal}{Journal of Physics: Condensed Matter}}
  \textbf{\bibinfo{volume}{16}}, \bibinfo{pages}{7439--7446}
  (\bibinfo{year}{2004}).

\bibitem{Zong2019}
\bibinfo{author}{Zong, A.} \emph{et~al.}
\newblock \bibinfo{title}{Evidence for topological defects in a photoinduced
  phase transition}.
\newblock \emph{\bibinfo{journal}{Nature Physics}}
  \textbf{\bibinfo{volume}{15}}, \bibinfo{pages}{27--31}
  (\bibinfo{year}{2019}).

\bibitem{Hsu2008}
\bibinfo{author}{Hsu, F.-C.} \emph{et~al.}
\newblock \bibinfo{title}{Superconductivity in the {PbO}-type structure
  {$\alpha$-FeSe}}.
\newblock \emph{\bibinfo{journal}{Proceedings of the National Academy of
  Sciences USA}} \textbf{\bibinfo{volume}{105}}, \bibinfo{pages}{14262--14264}
  (\bibinfo{year}{2008}).

\bibitem{Yumigeta2021}
\bibinfo{author}{Yumigeta, K.} \emph{et~al.}
\newblock \bibinfo{title}{Advances in rare-earth tritelluride quantum
  materials: Structure, properties, and synthesis}.
\newblock \emph{\bibinfo{journal}{Adv. Sci.}} \textbf{\bibinfo{volume}{8}},
  \bibinfo{pages}{2004762} (\bibinfo{year}{2021}).

\bibitem{Singh2024}
\bibinfo{author}{Singh, A.~G.} \emph{et~al.}
\newblock \bibinfo{title}{Emergent tetragonality in a fundamentally
  orthorhombic material}.
\newblock \emph{\bibinfo{journal}{Science Advances}}
  \textbf{\bibinfo{volume}{10}}, \bibinfo{pages}{eadk3321}
  (\bibinfo{year}{2024}).

\bibitem{LKAG1987}
\bibinfo{author}{Liechtenstein, A.~I.}, \bibinfo{author}{Katsnelson, M.~I.},
  \bibinfo{author}{Antropov, V.~P.} \& \bibinfo{author}{Gubanov, V.~A.}
\newblock \bibinfo{title}{Local spin density functional approach to the theory
  of exchange interactions in ferromagnetic metals and alloys}.
\newblock \emph{\bibinfo{journal}{J. Magn. Magn. Mater.}}
  \textbf{\bibinfo{volume}{67}}, \bibinfo{pages}{65 -- 74}
  (\bibinfo{year}{1987}).
\newblock
  \urlprefix\url{http://www.sciencedirect.com/science/article/pii/0304885387907219}.

\bibitem{Grechnev2007}
\bibinfo{author}{Grechnev, A.} \emph{et~al.}
\newblock \bibinfo{title}{Theory of bulk and surface quasiparticle spectra for
  {Fe}, {Co}, and {Ni}}.
\newblock \emph{\bibinfo{journal}{Phys. Rev. B}} \textbf{\bibinfo{volume}{76}},
  \bibinfo{pages}{035107} (\bibinfo{year}{2007}).

\bibitem{Sandratskii1992}
\bibinfo{author}{Sandratskii, L.~M.} \& \bibinfo{author}{K{\"u}bler, J.}
\newblock \bibinfo{title}{Static nonuniform magnetic susceptibility of selected
  transition metals}.
\newblock \emph{\bibinfo{journal}{J. Phys.: Condens. Matter}}
  \textbf{\bibinfo{volume}{4}}, \bibinfo{pages}{6927} (\bibinfo{year}{1992}).

\bibitem{Janak1977}
\bibinfo{author}{Janak, J.~F.}
\newblock \bibinfo{title}{Uniform susceptibilities of metallic elements}.
\newblock \emph{\bibinfo{journal}{Physical Review B}}
  \textbf{\bibinfo{volume}{16}}, \bibinfo{pages}{255--262}
  (\bibinfo{year}{1977}).

\bibitem{Gunnarsson1976}
\bibinfo{author}{Gunnarsson, O.}
\newblock \bibinfo{title}{Band model for magnetism of transition metals in the
  spin-density-functional formalism}.
\newblock \emph{\bibinfo{journal}{Journal of Physics F: Metal Physics}}
  \textbf{\bibinfo{volume}{6}}, \bibinfo{pages}{587} (\bibinfo{year}{1976}).
\newblock \urlprefix\url{https://doi.org/10.1088/0305-4608/6/4/018}.

\bibitem{Rajagopal1973}
\bibinfo{author}{Rajagopal, A.~K.}, \bibinfo{author}{Rath, J.} \&
  \bibinfo{author}{Kimball, J.~C.}
\newblock \bibinfo{title}{Spin correlations in a magnetic electron gas}.
\newblock \emph{\bibinfo{journal}{Phys. Rev. B}} \textbf{\bibinfo{volume}{7}},
  \bibinfo{pages}{2657} (\bibinfo{year}{1973}).

\bibitem{Gunnarsson1976LSD}
\bibinfo{author}{Gunnarsson, O.} \& \bibinfo{author}{Lundqvist, B.~I.}
\newblock \bibinfo{title}{Exchange and correlation in atoms, molecules, and
  solids by the spin-density-functional formalism}.
\newblock \emph{\bibinfo{journal}{Phys. Rev. B}} \textbf{\bibinfo{volume}{13}},
  \bibinfo{pages}{4274} (\bibinfo{year}{1976}).

\bibitem{Gunnarsson1977erratum}
\bibinfo{author}{Gunnarsson, O.} \& \bibinfo{author}{Lundqvist, B.~I.}
\newblock \bibinfo{title}{Erratum: Exchange and correlation in atoms,
  molecules, and solids by the spin-density-functional formalism}.
\newblock \emph{\bibinfo{journal}{Phys. Rev. B}} \textbf{\bibinfo{volume}{15}},
  \bibinfo{pages}{6006} (\bibinfo{year}{1977}).

\bibitem{oliver1979spin}
\bibinfo{author}{Oliver, G.} \& \bibinfo{author}{Perdew, J.}
\newblock \bibinfo{title}{Spin-density gradient expansion for the kinetic
  energy}.
\newblock \emph{\bibinfo{journal}{Physical Review A}}
  \textbf{\bibinfo{volume}{20}}, \bibinfo{pages}{397} (\bibinfo{year}{1979}).

\bibitem{Kumar2010}
\bibinfo{author}{Kumar, R.~S.} \emph{et~al.}
\newblock \bibinfo{title}{Crystal and electronic structure of {FeSe} at high
  pressure and low temperature}.
\newblock \emph{\bibinfo{journal}{J. Phys. Chem. B}}
  \textbf{\bibinfo{volume}{114}}, \bibinfo{pages}{12597--12606}
  (\bibinfo{year}{2010}).

\bibitem{Kiss2007}
\bibinfo{author}{Kiss, T.} \emph{et~al.}
\newblock \bibinfo{title}{Charge-order-maximized momentum-dependent
  superconductivity}.
\newblock \emph{\bibinfo{journal}{Nature Physics}}
  \textbf{\bibinfo{volume}{3}}, \bibinfo{pages}{720--725}
  (\bibinfo{year}{2007}).

\bibitem{Rossnagel2001}
\bibinfo{author}{Rossnagel, K.} \emph{et~al.}
\newblock \bibinfo{title}{{Fermi} surface of {2H-NbSe$_2$} and its implications
  on the charge-density-wave mechanism}.
\newblock \emph{\bibinfo{journal}{Physical Review B}}
  \textbf{\bibinfo{volume}{64}}, \bibinfo{pages}{235119}
  (\bibinfo{year}{2001}).

\bibitem{Straquadine2022}
\bibinfo{author}{Straquadine, J. A.~W.}, \bibinfo{author}{Ikeda, M.~S.} \&
  \bibinfo{author}{Fisher, I.~R.}
\newblock \bibinfo{title}{Evidence for realignment of the charge density wave
  state in {ErTe$_3$} and {TmTe$_3$} under uniaxial stress via elastocaloric
  and elastoresistivity measurements}.
\newblock \emph{\bibinfo{journal}{Phys. Rev. X}} \textbf{\bibinfo{volume}{12}},
  \bibinfo{pages}{021046} (\bibinfo{year}{2022}).

\bibitem{GalloFrantz2024}
\bibinfo{author}{Gallo-Frantz, A.} \emph{et~al.}
\newblock \bibinfo{title}{Charge density waves tuned by biaxial tensile
  stress}.
\newblock \emph{\bibinfo{journal}{Nat. Commun.}} \textbf{\bibinfo{volume}{15}},
  \bibinfo{pages}{3667} (\bibinfo{year}{2024}).

\bibitem{Singh2023thesis}
\bibinfo{author}{Singh, A.~G.}
\newblock \emph{\bibinfo{title}{Strain-Tuning Charge Density Wave Order in Rare
  Earth Tritellurides}}.
\newblock \bibinfo{type}{{PhD} thesis}, \bibinfo{school}{Stanford University}
  (\bibinfo{year}{2023}).
\newblock \urlprefix\url{https://purl.stanford.edu/kd339mm9275}.
\newblock \bibinfo{note}{Department of Applied Physics}.

\bibitem{Maschek2018}
\bibinfo{author}{Maschek, M.} \emph{et~al.}
\newblock \bibinfo{title}{Competing soft phonon modes at the
  charge-density-wave transitions in {DyTe$_3$}}.
\newblock \emph{\bibinfo{journal}{Phys. Rev. B}} \textbf{\bibinfo{volume}{98}},
  \bibinfo{pages}{094304} (\bibinfo{year}{2018}).

\bibitem{Yao2006}
\bibinfo{author}{Yao, H.}, \bibinfo{author}{Robertson, J.~A.},
  \bibinfo{author}{Kim, E.-A.} \& \bibinfo{author}{Kivelson, S.~A.}
\newblock \bibinfo{title}{Theory of stripes in quasi-two-dimensional rare-earth
  tellurides}.
\newblock \emph{\bibinfo{journal}{Phys. Rev. B}} \textbf{\bibinfo{volume}{74}},
  \bibinfo{pages}{245126} (\bibinfo{year}{2006}).

\bibitem{Kogar2020}
\bibinfo{author}{Kogar, A.} \emph{et~al.}
\newblock \bibinfo{title}{Light-induced charge density wave in {LaTe$_3$}}.
\newblock \emph{\bibinfo{journal}{Nat. Phys.}} \textbf{\bibinfo{volume}{16}},
  \bibinfo{pages}{159--163} (\bibinfo{year}{2020}).

\bibitem{Brouet2004}
\bibinfo{author}{Brouet, V.} \emph{et~al.}
\newblock \bibinfo{title}{Fermi surface reconstruction in the {CDW} state of
  {CeTe$_3$} observed by photoemission}.
\newblock \emph{\bibinfo{journal}{Physical Review Letters}}
  \textbf{\bibinfo{volume}{93}}, \bibinfo{pages}{126405}
  (\bibinfo{year}{2004}).

\bibitem{Lovesey1984}
\bibinfo{author}{Lovesey, S.~W.}
\newblock \emph{\bibinfo{title}{Theory of Neutron Scattering from Condensed
  Matter}} (\bibinfo{publisher}{Oxford University Press},
  \bibinfo{address}{Oxford}, \bibinfo{year}{1984}).

\bibitem{Taroni2011}
\bibinfo{author}{Taroni, A.}, \bibinfo{author}{Bergman, A.},
  \bibinfo{author}{Bergqvist, L.}, \bibinfo{author}{Hellsvik, J.} \&
  \bibinfo{author}{Eriksson, O.}
\newblock \bibinfo{title}{Suppression of standing spin waves in low-dimensional
  ferromagnets}.
\newblock \emph{\bibinfo{journal}{Phys. Rev. Lett.}}
  \textbf{\bibinfo{volume}{107}}, \bibinfo{pages}{037202}
  (\bibinfo{year}{2011}).

\bibitem{McQueen2009}
\bibinfo{author}{McQueen, T.~M.} \emph{et~al.}
\newblock \bibinfo{title}{Tetragonal-to-orthorhombic structural phase
  transition at 90~{K} in the superconductor {Fe$_{1.01}$Se}}.
\newblock \emph{\bibinfo{journal}{Physical Review Letters}}
  \textbf{\bibinfo{volume}{103}}, \bibinfo{pages}{057002}
  (\bibinfo{year}{2009}).

\bibitem{WatsonPRB2015}
\bibinfo{author}{Watson, M.~D.}, \bibinfo{author}{Haghighirad, A.~A.},
  \bibinfo{author}{Rhodes, L.~C.}, \bibinfo{author}{Hoesch, M.} \&
  \bibinfo{author}{Kim, T.~K.}
\newblock \bibinfo{title}{Emergence of the nematic electronic state in {FeSe}}.
\newblock \emph{\bibinfo{journal}{Physical Review B}}
  \textbf{\bibinfo{volume}{91}}, \bibinfo{pages}{155106}
  (\bibinfo{year}{2015}).

\bibitem{Maletz2014}
\bibinfo{author}{Maletz, J.} \emph{et~al.}
\newblock \bibinfo{title}{Unusual band renormalization in the simplest
  iron-based superconductor {FeSe$_{1-x}$}}.
\newblock \emph{\bibinfo{journal}{Physical Review B}}
  \textbf{\bibinfo{volume}{89}}, \bibinfo{pages}{220506}
  (\bibinfo{year}{2014}).

\end{thebibliography}


\begin{thebibliography}{1}
\expandafter\ifx\csname url\endcsname\relax
  \def\url#1{\texttt{#1}}\fi
\expandafter\ifx\csname urlprefix\endcsname\relax\def\urlprefix{URL }\fi
\providecommand{\bibinfo}[2]{#2}
\providecommand{\eprint}[2][]{\url{#2}}

\bibitem{cardias2017bethe}
\bibinfo{author}{Cardias, R.} \emph{et~al.}
\newblock \bibinfo{title}{The {Bethe}-{Slater} curve revisited; new insights
  from electronic structure theory}.
\newblock \emph{\bibinfo{journal}{Scientific reports}}
  \textbf{\bibinfo{volume}{7}}, \bibinfo{pages}{4058} (\bibinfo{year}{2017}).

\bibitem{Kumar2010}
\bibinfo{author}{Kumar, R.~S.} \emph{et~al.}
\newblock \bibinfo{title}{Crystal and electronic structure of {FeSe} at high
  pressure and low temperature}.
\newblock \emph{\bibinfo{journal}{J. Phys. Chem. B}}
  \textbf{\bibinfo{volume}{114}}, \bibinfo{pages}{12597--12606}
  (\bibinfo{year}{2010}).

\bibitem{Johannes2006}
\bibinfo{author}{Johannes, M.~D.}, \bibinfo{author}{Mazin, I.~I.} \&
  \bibinfo{author}{Howells, C.~A.}
\newblock \bibinfo{title}{Fermi-surface nesting and the origin of the
  charge-density wave in {NbSe$_2$}}.
\newblock \emph{\bibinfo{journal}{Physical Review B}}
  \textbf{\bibinfo{volume}{73}}, \bibinfo{pages}{205102}
  (\bibinfo{year}{2006}).

\end{thebibliography}

\newpage
\onecolumngrid
\appendix

\section{Calculated $U$ matrices}
\label{app:U}

Table~\ref{t:U_values} lists the exchange-correlation kernels calculated from Eq.~\eqref{eq:Ukernel}. The figures in the main text use the scanned values given with each figure (Sec.~\ref{subsec:compdet}).

\renewcommand{\arraystretch}{1.5}
\setlength{\tabcolsep}{5pt}
\begin{table*}
   \caption{Calculated exchange-correlation kernel $\mathbf{U}$ [Eq.~\eqref{eq:Ukernel}] per atom, in Ry, in the self-consistent PBE states used in this work. Each kernel is evaluated from Eq.~\eqref{eq:kernel_def}, with the valence spin density of the muffin-tin sphere of the listed atom as the fluctuation shape, and acts, in the RPA, on the listed orbitals of that atom, which are those that enter $\chi^0$ (Table~\ref{tab:comp}). In FeSe the Se~$4p$ orbitals also enter $\chi^0$ but carry no kernel. LaTe$_3$ has three inequivalent Te sites, Te1 in the LaTe layer and Te2 and Te3 in the square nets; the kernels of Te2 and Te3 agree to $10^{-5}$~Ry. For the magnetic systems the values are for a spin-up site; on spin-down sites $U_{\uparrow\uparrow}$ and $U_{\downarrow\downarrow}$ are interchanged. bcc~Cr is the G-type antiferromagnet at $a=5.29$~bohr ($0.62~\mu_\mathrm{B}$) and FeSe the single-stripe antiferromagnet ($2.52~\mu_\mathrm{B}$ per Fe). In all systems $U_{\uparrow\downarrow}$ is 7--10\% of the mean diagonal element, which supports $U_{\uparrow\downarrow}=0.1\,U_{\uparrow\uparrow}$ in the scans. The last column gives the Stoner parameter $I=\tfrac12(U_{\uparrow\uparrow}+U_{\downarrow\downarrow})-U_{\uparrow\downarrow}$ of Eq.~\eqref{eq:stoner}, the quantity used in Sec.~\ref{subsec:compdet}. For the nonmagnetic states of Cr and FeSe we use the kernel of the antiferromagnetic state; for Cr, the kernel of the nonmagnetic state at $a=5.45$~bohr gives $I=0.0788$~Ry, about $1\%$ below the value listed. In the magnetic states $U_{\downarrow\downarrow}$ exceeds $U_{\uparrow\uparrow}$ by 7\% (Cr), 19\% (Fe) and 25\% (FeSe), so taking $U_{\downarrow\downarrow}=U_{\uparrow\uparrow}$ there remains an approximation.}
  \centering
  \vspace{5pt}
  \begin{tabular}{c|c|c|c|cccc}
    \hline
    Density wave & Compound & Atom & Orbitals & $U_{\uparrow\uparrow}$ & $U_{\downarrow\downarrow}$ & $U_{\uparrow\downarrow} = U_{\downarrow\uparrow}$ & $I$ \\
    \hline
      $-$ & bcc Fe (FM)          & Fe                     & Fe $3d$             & $0.0782$ & $0.0930$ & $0.0061$ & $0.0795$ \\
      SDW & bcc Cr (AFM)         & Cr                     & Cr $4s$, $4p$, $3d$ & $0.0838$ & $0.0895$ & $0.0070$ & $0.0797$ \\
      SDW & FeSe (AFM)           & Fe                     & Fe $3d$             & $0.0783$ & $0.0979$ & $0.0062$ & $0.0819$ \\
      CDW & 2H-NbSe$_2$          & Nb                     & Nb $4d$             & $0.0751$ & $0.0751$ & $0.0075$ & $0.0676$ \\
      CDW & LaTe$_3$             & Te1 (LaTe layer)       & Te $5p$             & $0.0475$ & $0.0475$ & $0.0041$ & $0.0434$ \\
      CDW & LaTe$_3$             & Te2, Te3 (square nets) & Te $5p$             & $0.0484$ & $0.0484$ & $0.0041$ & $0.0442$ \\
    \hline
  \end{tabular}
  \label{t:U_values}
\end{table*}

We have used an energy-based definition of the exchange-correlation kernel, since it provides $U$-matrices that are symmetric for both LDA and GGA functionals. The kernel is the second derivative of the muffin-tin exchange--correlation energy with respect to the number of electrons of each spin, for a fixed fluctuation
shape $g_\sigma$ normalized as $\int_{\mathrm{MT}} g_\sigma\, d^3r = 1$. The fluctuation shape $g_\sigma(\mathbf{r})$ describes how the electron density is varied at each point of the muffin tin. Only the valence density is varied, while the core density is fixed, since it is not involved in the susceptibility analysis.

We have considered two fluctuation shapes: the valence spin density itself, and the squared radial $d$-orbital density at the Fermi energy, calculated from the radial Schr\"odinger equation in the spin-averaged DFT potential. The latter follows the definition in Ref.~\onlinecite{Gunnarsson1976}.

For the valence-density shape,
\begin{equation}
U_{\sigma\sigma'} \;=\; -\,
\frac{\partial^2 E_{xc}^{\mathrm{MT}}\!\left[\,n + \delta N_\uparrow\, g_\uparrow
      + \delta N_\downarrow\, g_\downarrow\right]}
     {\partial N_\sigma\,\partial N_{\sigma'}}\Bigg|_{\delta N = 0},
\qquad
g_\sigma(\mathbf r) = \frac{n^{v}_\sigma(\mathbf r)}{N^{v}_\sigma},\quad
N^{v}_\sigma = \int_{\mathrm{MT}} n^{v}_\sigma\, d^3r .
\label{eq:kernel_def}
\end{equation}
Here $n^v_\sigma$ ($\sigma = \uparrow, \downarrow$) is the valence spin density inside the muffin-tin sphere, and $E_{xc}^{\mathrm{MT}} = \int_{\mathrm{MT}} n\,\varepsilon_{xc}\,d^3r$. The minus sign is that of Eq.~\eqref{eq:Ukernel}, so that the elements of $\mathbf{U}$ are positive.

\subsection{LDA form}

For a local functional like LDA, Eq.~\eqref{eq:kernel_def} reduces to the shape-weighted average of the local kernel in Eq.~(14) of Ref.~\onlinecite{katsnelson2004magnetic}:
\begin{equation}
U_{\sigma\sigma'} = -\int_{\mathrm{MT}} d^3r\; g_\sigma(\mathbf r)\,
\frac{\partial^2 (n\varepsilon_{xc})}{\partial n_\sigma\,\partial n_{\sigma'}}\,
g_{\sigma'}(\mathbf r) .
\label{eq:kernel_lda}
\end{equation}
For a non-local functional like GGA, this expression is not applicable, and Eq.~\eqref{eq:kernel_def} should be used.

\subsection{$d$-orbital fluctuation shape}

The second fluctuation shape is based on the squared radial $d$ orbital at the Fermi energy, as in Eq.~(2.8) of Ref.~\onlinecite{Gunnarsson1976}:
\begin{equation}
g^{d}_\uparrow(\mathbf r) = g^{d}_\downarrow(\mathbf r)
  = \frac{\phi_{d}^2(r,E_F)}{4\pi},
\qquad \mathrm{with\ normalization}\quad
\int_0^{S_{\mathrm{MT}}} \phi_{d}^2(r,E_F)\, r^2\, dr = 1 .
\label{eq:dorbital_shape}
\end{equation}
Here $\phi_{d}(r,E)$ is the regular $l=2$ solution of the scalar-relativistic radial Schr\"odinger equation in the spin-averaged spherical potential $\tfrac12(V_\uparrow + V_\downarrow)$. Because the shape defined by Eq.~\eqref{eq:dorbital_shape} is spherical, the LDA form \eqref{eq:kernel_lda} becomes a radial integral over the spherically averaged $\bar U_{\sigma\sigma'}(r)$:
\begin{equation}
U^{d}_{\sigma\sigma'} = -\frac{1}{4\pi}\int_0^{S_{\mathrm{MT}}} r^2\, dr\;
\phi_{d}^4(r,E_F)\,\bar U_{\sigma\sigma'}(r),
\qquad
\bar U_{\sigma\sigma'}(r) = \frac{1}{4\pi}\int d\Omega\,
\frac{\partial^2 (n\varepsilon_{xc})}{\partial n_\sigma\,\partial n_{\sigma'}} .
\label{eq:kernel_d_lda}
\end{equation}

\textit{Stoner parameter.} For both shapes we also evaluate the Stoner parameter $I=\tfrac12(U_{\uparrow\uparrow}+U_{\downarrow\downarrow})-U_{\uparrow\downarrow}$ of Eq.~\eqref{eq:stoner}. With the $d$-orbital shape this corresponds exactly to Eq.~(2.15) in Ref.~\onlinecite{Gunnarsson1976} with the Wigner--Seitz radius replaced by the muffin-tin
radius. To show this, note that $I = -\partial(v^{xc}_{+} - v^{xc}_{-})/\partial m $ at fixed $n$. With $\phi_d = \phi_2$, in the limit $m \rightarrow 0$ and with a local spin density potential [Eqs.~(2.8) and (2.13) in Ref.~\onlinecite{Gunnarsson1976}], one can obtain $v^{xc}_{+} - v^{xc}_{-} \simeq \tfrac23\,\mu^x\delta\, m/\rho$, so
Eqs.~\eqref{eq:kernel_d_lda} and \eqref{eq:stoner} give
\begin{equation}
I = -\int_0^{S_{\mathrm{MT}}} r^2\, dr
    \left(\frac{\mu^x(r_s)}{6\pi}\right)\delta(r_s)\,
    \frac{\phi_2^2(r,E_F)\,\phi_2^2(r,E_F)}{\rho(r)} ,
\label{eq:gunnarsson}
\end{equation}

We evaluate the Stoner parameter using Eq.~\eqref{eq:stoner} for the spin-polarized state with the full exchange-correlation functional but using the orbital $\phi_{d}(r,E_\mathrm{F})$ taken from the spin-averaged potential, as explained above.

\subsection{Numerical evaluation}

Central differences are used for numerical calculation of the exchange-correlation kernel [Eq.~\eqref{eq:kernel_def}], with $n_\sigma \to n_\sigma + \delta N_\sigma\,
g_\sigma$ and $\delta N_\sigma = \epsilon N^v_\sigma$ for both shapes (for the
valence shape this is $n_\sigma \to n_\sigma + \epsilon\, n^v_\sigma$):
\begin{equation}
U_{\sigma\sigma} \simeq
-\frac{E(+\delta N_\sigma) - 2E(0) + E(-\delta N_\sigma)}{(\delta N_\sigma)^2},
\qquad
U_{\uparrow\downarrow} \simeq
-\frac{E(+,+) - E(+,-) - E(-,+) + E(-,-)}{4\,\delta N_\uparrow\, \delta N_\downarrow},
\label{eq:fd}
\end{equation}
with $\epsilon = 0.005$. For the valence shape, the results change by less than
about $10^{-6}$ (relative) for $\epsilon$ between $2\times 10^{-4}$ and
$2\times 10^{-2}$. Only the muffin-tin spheres contribute to these expressions; the interstitial region is not included. The values in Table~\ref{t:U_values} were obtained using the valence-density shape. As a test, for ferromagnetic bcc~Fe (PBE, $2.12~\mu_\mathrm{B}$) the two shapes give the Stoner parameter $I=1.08$~eV (valence density) and $1.05$~eV ($d$ orbital), and with LDA the $d$-orbital value, $0.97$~eV, is close to $0.92$~eV reported in Table~1 of Ref.~\onlinecite{Gunnarsson1976}.

\end{document}


\title{Supplemental Material for ``A Real-Space Framework for the Bare and Interacting Longitudinal Susceptibility of Solids''}

\author{Rohit Pathak}
    \affiliation{\Uppsala}
    \affiliation{\WISEUppsala}

\author{Abhay Mishra}
    \affiliation{\Uppsala}

\author{Mikhail I. Katsnelson}
    \affiliation{\WISEUppsala}
    \affiliation{\Netherlands}

\author{Anna Delin}
    \affiliation{\KTH}
    \affiliation{\WISEKTH}
    \affiliation{\SeRC}

\author{Olle Eriksson}
    \affiliation{\Uppsala}
    \affiliation{\WISEUppsala}

\author{Vladislav Borisov}
    \affiliation{\Uppsala}
    \affiliation{\WISEUppsala}

\date{\today}

\maketitle
\raggedbottom
\section{Bond vectors and dependence on the unit cell choice}

\begin{figure}[H]
  \centering
  \includegraphics[width=\columnwidth]{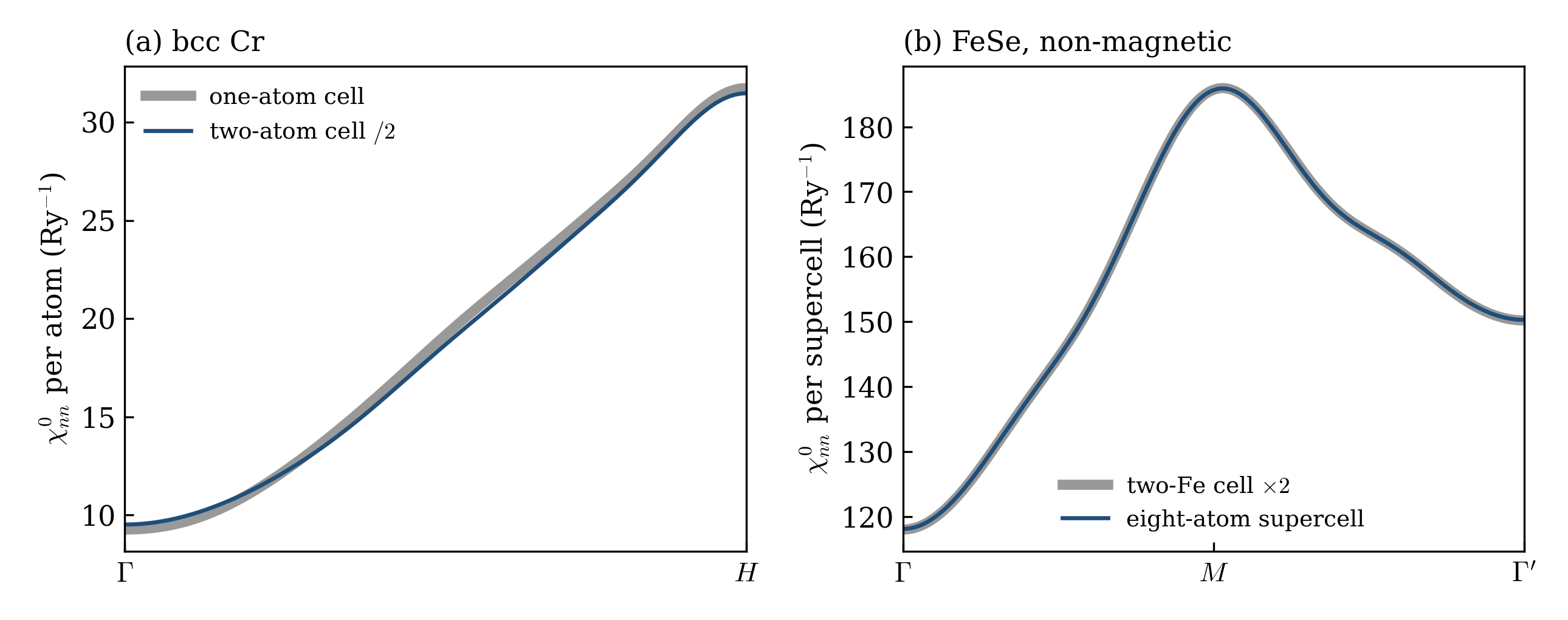}
  \caption{The susceptibility calculated using two different unit cells of the same system. (a) bcc~Cr, one-atom and two-atom cells, per atom. (b) Nonmagnetic FeSe, two-Fe cell ($\chi_{nn}^0$ is multiplied by two here)
  and four-Fe supercell.}
  \label{fig:cells}
\end{figure}

We obtain the bond vectors of Eq.~(9) of the main text by
requiring that the bonds of every site form one complete sphere of radius
$r_\mathrm{cut}$. Figure~\ref{fig:cells} checks the result on two pairs of
cells that describe the same crystal: bcc~Cr in the one- and two-atom cells
agrees to within $0.5~\mathrm{Ry}^{-1}$ per atom along $\Gamma$-$H$, and
nonmagnetic FeSe in the eight-atom supercell reproduces twice the two-Fe
cell to within $0.1~\mathrm{Ry}^{-1}$, on a variation of $68~\mathrm{Ry}^{-1}$.

\section{Convergence of the real-space sum}

\begin{figure}[H]
  \centering
  \includegraphics[width=\columnwidth]{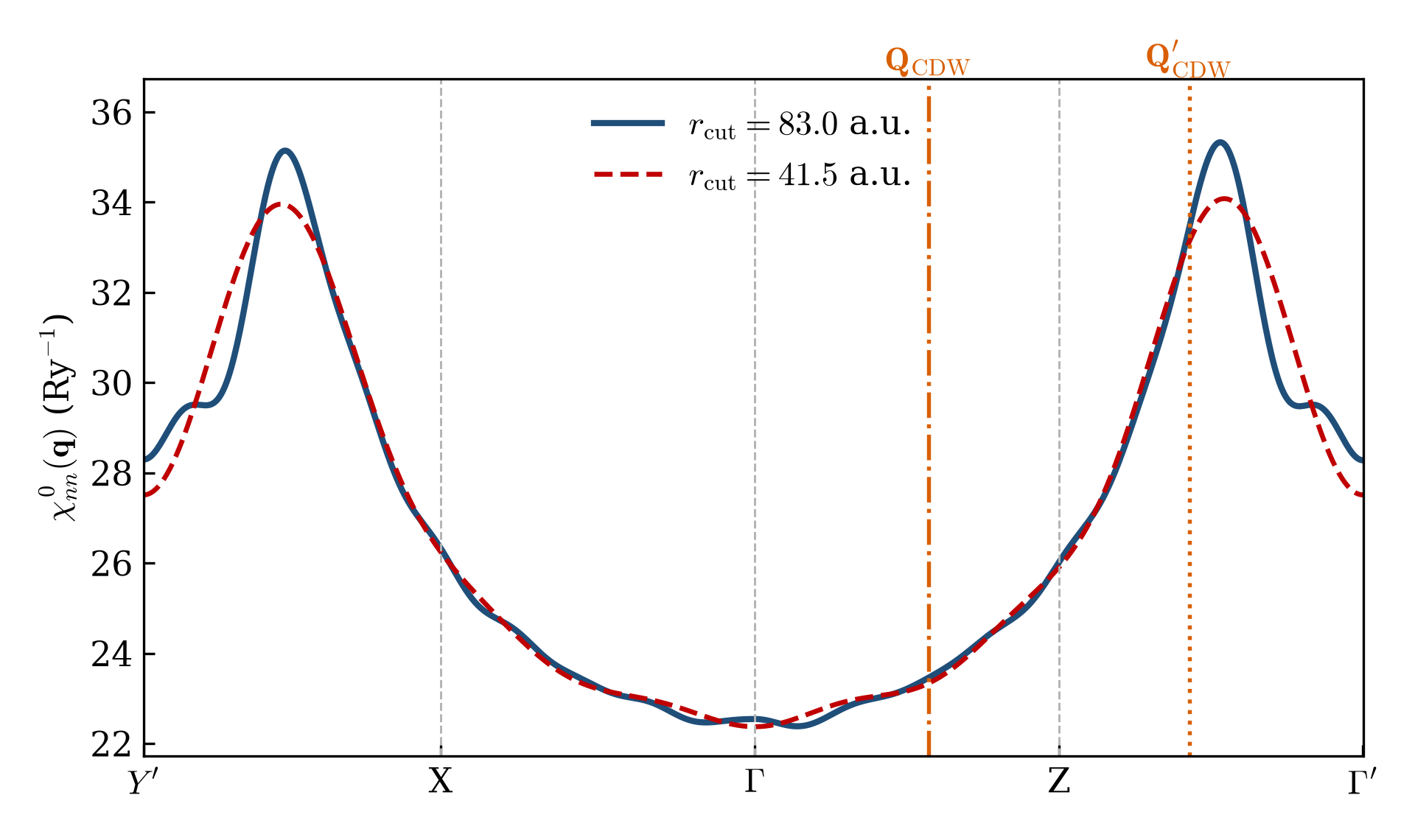}
    \caption{Bare charge susceptibility of LaTe$_3$ along
  $Y'$-$X$-$\Gamma$-$Z$-$\Gamma'$ for $r_\mathrm{cut}=83.0$~a.u.\ (solid) and
  $41.5$~a.u.\ (dashed); the shorter cutoff keeps only the bonds within
  $41.5$~a.u.\ of the same calculation. Halving $r_\mathrm{cut}$ changes
  $\chi^0_{nn}$ by $1.2\%$ on average and at most $6\%$, and moves the maximum by less than $2\%$ of the $\Gamma$-$Z$ distance.}
  \label{fig:rcut}
\end{figure}

The bare susceptibility $\chi^0(\mathbf{q})$ is obtained from Eq.~(9) of the main text, a sum of the real-space bubbles $\chi^0_{\sigma,ij}(\mathbf{R})$ over bonds. Only bonds of length $|\mathbf{d}_{ij}|\le r_\mathrm{cut}$ are included. Table~\ref{tab:rcut} lists $r_\mathrm{cut}$ for each system.
Figure~\ref{fig:rcut} shows the effect of this cutoff on the susceptibility of LaTe$_3$, the material with the longest range of the
real-space response (Fig.~\ref{fig:chiR}): halving
$r_\mathrm{cut}$ changes the curve by about $1\%$ on average and leaves its maximum virtually unchanged.

\begin{table}[H]
\caption{Real-space cutoff $r_\mathrm{cut}$ for each system, in units of the lattice parameter $a$, in a.u.\ and in \AA. The values in a.u.\ are those of Table~II of the main text. For bcc~Cr in the one-atom cell, $r_\mathrm{cut}$ is seven nearest-neighbour distances.}
\label{tab:rcut}
\begin{ruledtabular}
\begin{tabular}{lcccc}
 System & $a$ (a.u.) & $r_\mathrm{cut}/a$ & $r_{\mathrm{cut}}$ (a.u.) & $r_{\mathrm{cut}}$ (\AA) \\ \hline
 2H-NbSe$_2$ & 6.5007 & 5    & 32.5 & 17.2 \\
 LaTe$_3$    & 8.3001 & 10   & 83.0 & 43.9 \\
 bcc Cr      & 5.4500 & 6.06 & 33.0 & 17.5 \\
 FeSe        & 7.1214 & 5    & 35.6 & 18.8 \\
 bcc Fe      & 5.27   & 7    & 36.9 & 19.5 \\
 bcc Cr, two-atom cells of Sec.~V~C & 5.2899 & 7 & 37.0 & 19.6 \\
\end{tabular}
\end{ruledtabular}
\end{table}

\section{Electronic temperature}

All results in the main text use a thermal smearing width $W=0.005$~Ry
($T\approx790$~K). To test the sensitivity to this choice we repeated the
bcc~Cr calculation at $W=0.005$, $0.002$ and $0.001$~Ry; the other systems
were not tested.

\begin{table}[H]
\caption{Peak position and height of $\chi^0_{nn}(\mathbf{q})$ in bcc~Cr as a
function of the smearing width.}
\label{tab:wscan}
\begin{ruledtabular}
\begin{tabular}{lccc}
 $W$ (Ry) & $T$ (K) & $\chi^0_{nn}(H)$ (Ry$^{-1}$) & peak position \\ \hline
 0.005 & 790 & 31.7 & $H$ \\
 0.002 & 316 & 32.3  & $H$ \\
 0.001 & 158 & 32.2  & $H$ \\
\end{tabular}
\end{ruledtabular}
\end{table}

\section{Sum rule}

At $\mathbf{q}=\mathbf{0}$ the bare charge susceptibility would equal, for a
complete orthonormal set of local orbitals, the
thermally averaged density of states of the retained orbitals,
\begin{equation}
  \chi^0_{nn}(\mathbf{0}) = \int \! d\varepsilon \, N(\varepsilon)
  \left( -\frac{\partial f}{\partial \varepsilon} \right),
  \label{eq:sumruleS}
\end{equation}
which reduces to $N(\varepsilon_F)$ at $T=0$.

Table~\ref{tab:sumrule} compares the two quantities for each system; they
agree to within $13\%$.

\begin{table}[H]
\caption{Numerical verification of the sum rule that the susceptibility $\chi^0_{nn}(\mathbf{0})$ should be equal to $\langle N\rangle$, the projected density of states of the retained orbitals convolved with $-\partial f/\partial \varepsilon$ at $W=0.005$~Ry. Deviation from this rule is listed for each studied system in the rightmost column.}
\label{tab:sumrule}
\begin{ruledtabular}
\begin{tabular}{lccc}
 System & $\chi^0_{nn}(\mathbf{0})$ & $\langle N\rangle$ & deviation \\ \hline
 FeSe        & 59.0 & 59.3 & $-0.5\%$ \\
  LaTe$_3$    & 22.5 & 20.7 & $+8.7\%$ \\
 2H-NbSe$_2$ & 28.0 & 31.2 & $-10.2\%$ \\
 bcc Cr      &  9.3 & 10.6 & $-13.0\%$ \\
\end{tabular}
\end{ruledtabular}
\end{table}

We attribute these deviations to the projection onto the muffin-tin region, meaning that the interstitial contribution to Eq.~\eqref{eq:sumruleS} is not included, and to the projected orbitals not being orthonormal.

\section{Real-space bare susceptibility}

\begin{figure}[H]
  \centering
  \includegraphics[width=\textwidth]{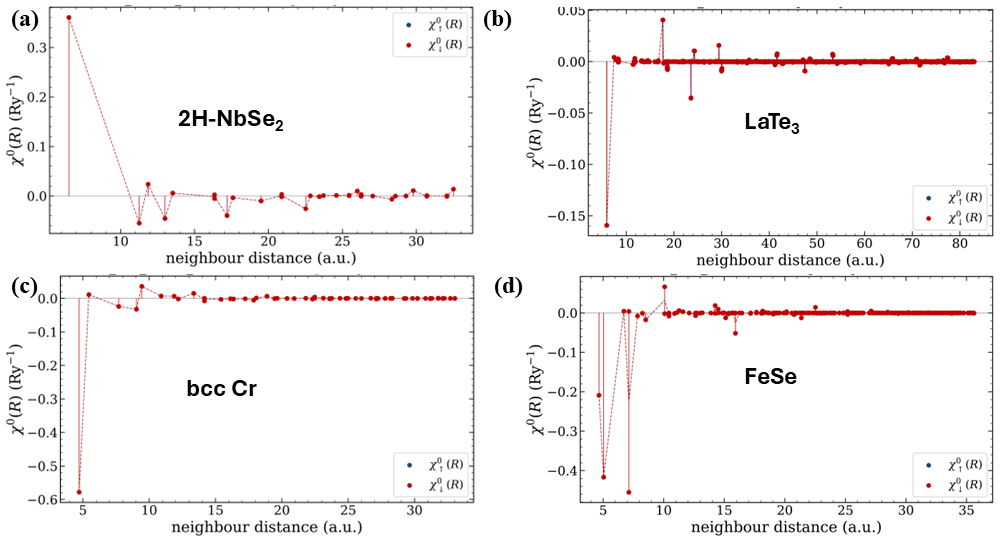}
  \caption{Intersite bare susceptibility $\chi^0_\sigma(\mathbf R)$ per
  spin, summed over the retained orbitals, for every bond within
  $r_\mathrm{cut}$, plotted as a function of the bond length: (a) 2H-NbSe$_2$ (Nb--Nb bonds),
  (b) LaTe$_3$ (Te--Te bonds), (c) bcc~Cr (one-atom cell), (d) FeSe (two-Fe cell;
  Fe--Fe, Fe--Se and Se--Se bonds). The two spin channels coincide in these
  nonmagnetic calculations. The on-site term ($\mathbf R=0$) is not shown.}
  \label{fig:chiR}
\end{figure}

Figure~\ref{fig:chiR} shows the intersite bubbles $\chi^0_\sigma(\mathbf R)$
of Eq.~(1) of the main text, summed over the retained orbitals, with one
point for every bond within $r_\mathrm{cut}$, against the bond length. In all
four materials the first few shells dominate, and contributions of both
signs occur. In bcc~Cr the nearest-neighbour $\langle111\rangle$ bonds
dominate, and the second-neighbour $\langle100\rangle$ bonds have the opposite
sign. At $H$ both shells add, and only the $\langle111\rangle$ shell
distinguishes $H$ from $\Gamma$ (Sec.~VI of the main text). In FeSe, the contributions of Fe--Fe bonds at distances $a/\sqrt2$ and $a$ (5.04 and
7.12~a.u.) are both large and negative, and the contribution of the Fe--Se bond at 4.65~a.u.\ is
about half as large. In 2H-NbSe$_2$ the in-plane nearest-neighbour bond shows a positive susceptibility, while further shells contribute much less. In LaTe$_3$ sizeable susceptibility is observed for many Te--Te bonds up to a distance of 23.5~a.u., which is why a larger cutoff is used for this material (Table~\ref{tab:rcut}). The orbital content
of the strongest shells is resolved in Sec.~\ref{sec:orbS}.

\section{Orbital decomposition of the bare susceptibility}
\label{sec:orbS}

\begin{figure}[H]
  \centering
  \includegraphics[width=0.95\columnwidth]{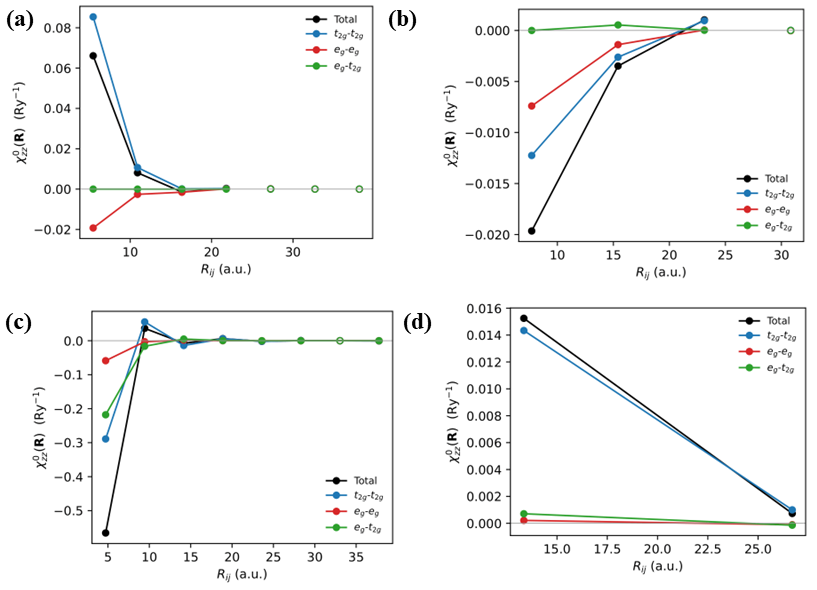}
  \caption{Distance dependence of the $t_{2g}$--$t_{2g}$, $e_g$--$e_g$ and
  $e_g$--$t_{2g}$ cross contributions to $\chi^0_{zz}(\mathbf{R})$ in bcc~Cr,
  resolved by bond direction: (a) $\langle 100\rangle$, (b) $\langle
  110\rangle$, (c) $\langle 111\rangle$ and (d) $\langle 112\rangle$. The
  three types of orbital-resolved contributions sum up to the total value (black lines) for every interaction shell. The $e_g$--$t_{2g}$ cross
  term vanishes exactly along $\langle 100\rangle$ due to symmetry considerations, is finite along
  $\langle 111\rangle$ and $\langle 112\rangle$ directions, and is allowed by symmetry but negligible along the $\langle 110\rangle$ direction.}
  \label{fig:t2geg}
\end{figure}
Figure~\ref{fig:t2geg} splits the real-space representation of the susceptibility of bcc~Cr into its $t_{2g}$--$t_{2g}$, $e_g$--$e_g$ and $e_g$--$t_{2g}$ blocks, with $e_g=\{d_{z^2},d_{x^2-y^2}\}$ and $t_{2g}=\{d_{yz},d_{xz},d_{xy}\}$, for bonds along four directions. Each block sum is the same for every bond of a shell, because the symmetry operations of a cubic lattice do not mix $e_g$ with $t_{2g}$ orbitals. The nearest-neighbour contribution is the largest and goes along $\langle111\rangle$. Here the $t_{2g}$ block gives about half of the total susceptibility ($-0.29$ of $-0.56~\mathrm{Ry}^{-1}$), the cross term contributes $-0.22~\mathrm{Ry}^{-1}$ and the $e_g$ block contributes $-0.06~\mathrm{Ry}^{-1}$. All three blocks are negative on this shell and add up. The phase factor of the nearest-neighbour bonds is $-1$ at $H$ and $+1$ at $\Gamma$, so each block on its own raises $\chi^0_{zz}(\mathbf{q})$ at $H$ relative to $\Gamma$. Each symmetry channel therefore favours the antiferromagnetic order of bcc~Cr and the spin-density wave, whose wave vector lies close to $H$. The orbital-resolved Heisenberg exchange of bcc~Cr behaves in the same way, with all contributions to the nearest-neighbour coupling of the same, antiferromagnetic sign, whereas in bcc Mn and Fe the $t_{2g}$--$t_{2g}$ contribution has the opposite sign to the other two~\cite{cardias2017bethe}. For the second-nearest-neighbor contribution, the values are smaller due to weaker wavefunction overlap. Along 
the $\langle100\rangle$ direction, a cross term vanishes due to symmetry reasons (see~\cite{cardias2017bethe}) and the $t_{2g}$ block 
($+0.09~\mathrm{Ry}^{-1}$) outweighs the $e_g$ block ($-0.02~\mathrm{Ry}^{-1}$) giving a total susceptibility that is positive.
These orbital contributions are calculated for the Cr~$3d$ states in the two-atom cubic cell
(Fig.~\ref{fig:mat_cr}). Also, due to weaker wavefunction overlap, the susceptibility along the $\langle110\rangle$ and $\langle112\rangle$ directions is smaller in all symmetry channels, with a strong dominance in the $t_{2g}$--$t_{2g}$ channel for the $\langle112\rangle$ direction. The example of Figure~\ref{fig:t2geg} demonstrates that a real-space formulation and calculation of the susceptibility allows for fundamentally deeper, symmetry-resolved analysis, as opposed to a pure reciprocal-space analysis.

\subsection{Orbital-resolved susceptibility on individual bonds}

\begin{figure}[H]
  \centering
  \includegraphics[width=\textwidth]{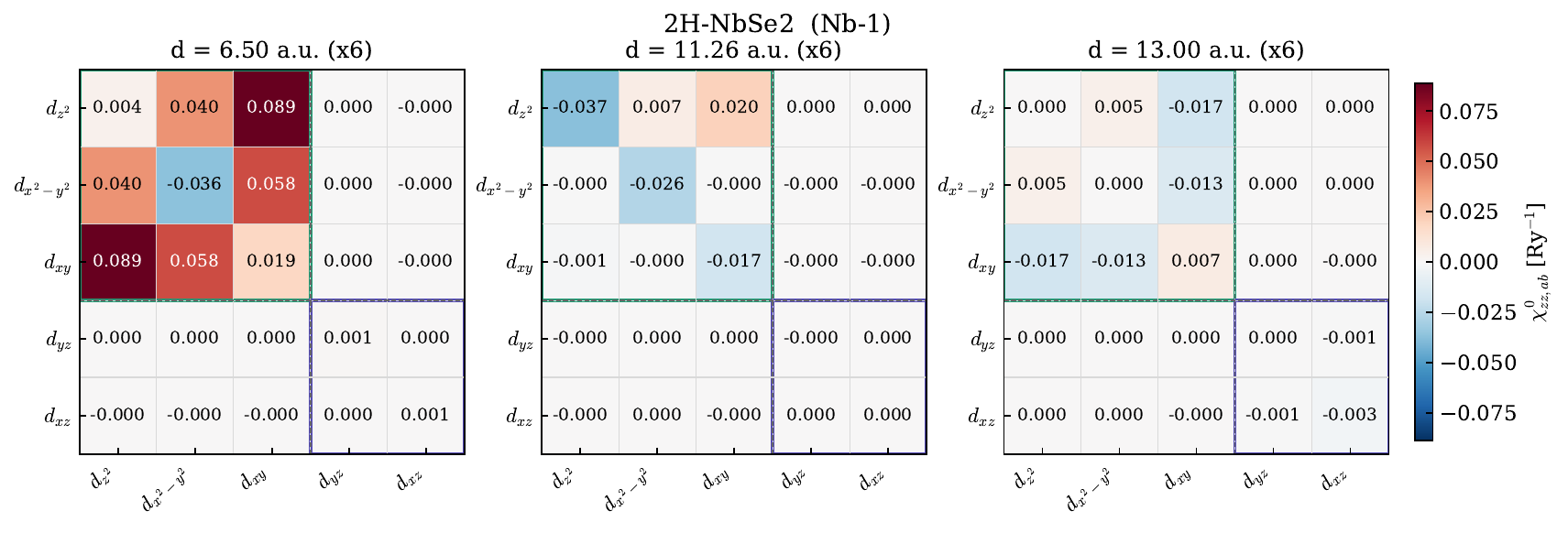}
  \caption{2H-NbSe$_2$: $\chi^0_{zz,ab}(\mathbf{R})$ (Ry$^{-1}$, per unit
  cell) between the Nb~$4d$ orbitals on one bond of each of the three
  strongest shells, central site Nb-1. Rows: orbital on the central site;
  columns: orbital on the neighbour. Boxes: orbitals even
  ($d_{z^2},d_{x^2-y^2},d_{xy}$) and odd ($d_{yz},d_{xz}$) under the mirror
  plane through the Nb layer.}
  \label{fig:mat_nbse2}
\end{figure}

\begin{figure}[H]
  \centering
  \includegraphics[width=\textwidth]{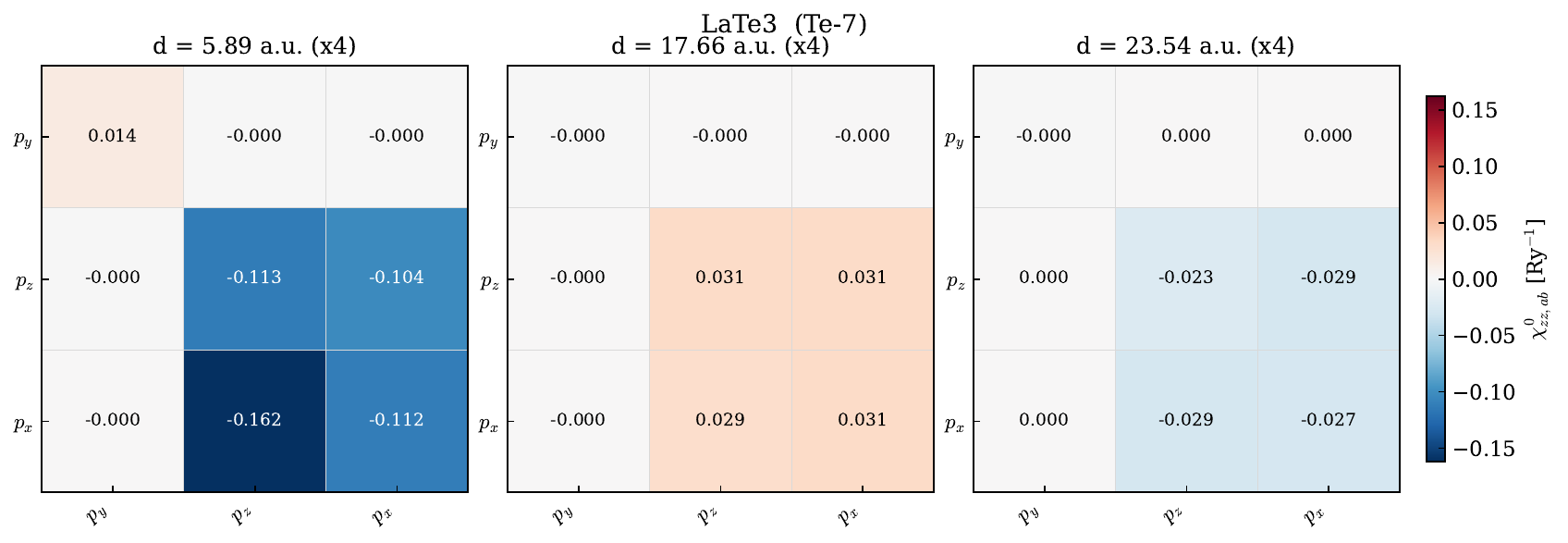}
  \caption{LaTe$_3$: as Fig.~\ref{fig:mat_nbse2}, for the Te~$5p$ orbitals
  of the square-net site Te-7. $y$ is the stacking axis, perpendicular to
  the net.}
  \label{fig:mat_late3}
\end{figure}

\begin{figure}[H]
  \centering
  \includegraphics[width=\textwidth]{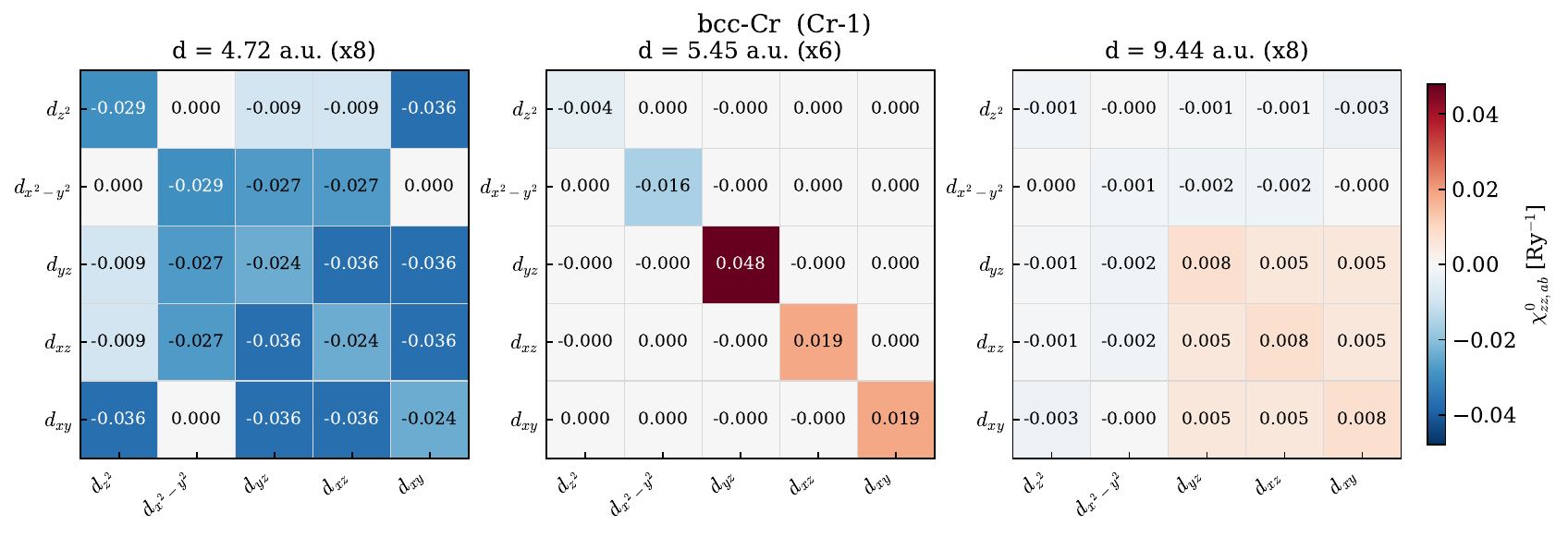}
  \caption{bcc~Cr: as Fig.~\ref{fig:mat_nbse2}, for the Cr~$3d$ orbitals of
  site Cr-1 in the two-atom cubic cell; axes along the cube edges.}
  \label{fig:mat_cr}
\end{figure}

\begin{figure}[H]
  \centering
  \includegraphics[width=\textwidth]{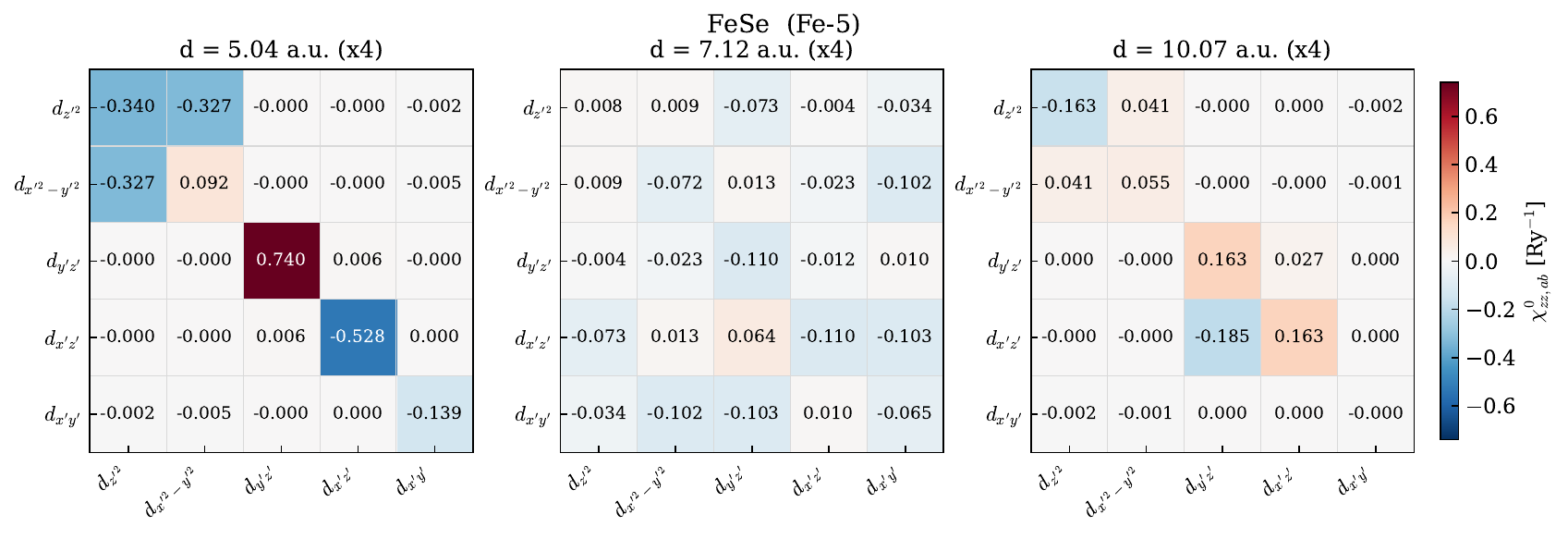}
  \caption{FeSe: as Fig.~\ref{fig:mat_nbse2}, for the Fe~$3d$ orbitals of
  site Fe-5 in the nonmagnetic $\sqrt2a\times\sqrt2a\times c$ cell. Primed
  labels refer to the local axes of the Fe site, with $z'$ along an in-plane nearest-neighbour Fe--Fe
  bond.}
  \label{fig:mat_fese}
\end{figure}

Figures~\ref{fig:mat_nbse2}--\ref{fig:mat_fese} give the full orbital
matrices behind the orbital table of the main text. For a central site $i$
and a neighbour $j$ at bond vector $\mathbf{R}$, the element
$\chi^0_{zz,ab}(\mathbf{R})$ is the contribution to the bare longitudinal
spin susceptibility of orbital $a$ on site $i$ and orbital $b$ on site $j$.
Rows label $a$ and columns label $b$. The orbitals are real harmonics:
$d_{z^2}, d_{x^2-y^2}, d_{yz}, d_{xz}, d_{xy}$ for the $d$ shells and
$p_x, p_y, p_z$ for the Te $5p$ shell. The values of $\chi^0_{zz,ab}(\mathbf{R})$ are in units of Ry$^{-1}$ and are
normalised, like the curves of the main text, as contributions per unit cell:
the per-spin response of one site multiplied by $N_s/2$, where $N_s$ is the
number of sites in the calculation ($N_s=2$, 6, 2 and 4 for NbSe$_2$,
LaTe$_3$, Cr and FeSe). Positive elements are red and negative ones blue,
and each element is printed in its cell.

The sum of all elements of a matrix is the contribution of that bond to
$\chi^0_{zz}$ in Eq.~(9) of the main text. Diagonal elements are
contributions of the same type of orbital at each end of the bond; off-diagonal elements
are contributions of two different orbitals. Bonds of equal length that are
related by symmetry form a shell; the panel title gives the bond length $d$
and the number of bonds in the shell. Each panel shows one bond of the
shell. For each material we show the three shells with the largest
$|\sum_{ab}\chi^0_{zz,ab}(\mathbf{R})|$, in order of increasing length.

The element sum is the same for every bond of a shell and independent
of the orientation of the orbital axes; individual elements depend on which
bond of the shell is shown and on the axes. The matrix of a bond need not be
symmetric, since $\chi^0_{ab}(\mathbf{R})=\chi^0_{ba}(-\mathbf{R})$. The
orbital axes are the Cartesian axes of the crystal for NbSe$_2$ ($z$ along
$c$) and Cr (the cube axes). For LaTe$_3$ they are the local axes of the Te
site, with $y$ along the stacking axis $b$, perpendicular to the Te nets, and
$x$ and $z$ in the plane of the nets; for FeSe they are the local axes of the
Fe site (primed), with $z'$ along an in-plane nearest-neighbour Fe--Fe bond and $x'$, $y'$ at $45^\circ$ to $c$.

\paragraph{2H-NbSe$_2$ (Fig.~\ref{fig:mat_nbse2}).} The three strongest
shells are in-plane Nb--Nb bonds at $a$, $\sqrt3\,a$ and $2a$ (6.50, 11.26
and 13.00~a.u., six bonds each). Only the in-plane orbitals $d_{z^2}$,
$d_{x^2-y^2}$ and $d_{xy}$ contribute; the boxes separate them from
$d_{yz}$ and $d_{xz}$, which are odd under the mirror plane through the Nb
layer, and the elements between the two blocks vanish on these bonds. On the
nearest-neighbour bond the largest elements are off-diagonal, between
$d_{z^2}$ and $d_{xy}$ and between $d_{x^2-y^2}$ and $d_{xy}$, so the
response mixes the in-plane orbitals.

\paragraph{LaTe$_3$ (Fig.~\ref{fig:mat_late3}).} The site is a Te atom of
the square net. The three strongest shells lie in the plane of the net, at
one, three and four Te--Te nearest-neighbour distances (5.89, 17.66 and
23.54~a.u., four bonds each), that is, along the Te--Te bonds of the net. The
orbital $p_y$, perpendicular to the net, is nearly decoupled from $p_x$ and $p_z$.
The response is carried by the in-plane orbitals $p_x$ and $p_z$, with
diagonal and off-diagonal elements of similar size, and its sign alternates
between the shells shown.

\paragraph{bcc Cr (Fig.~\ref{fig:mat_cr}).} The site is Cr in the two-atom
cubic cell. The three strongest shells are the nearest neighbours along
$\langle111\rangle$ (4.72~a.u., eight bonds), the second neighbours along
$\langle100\rangle$ (5.45~a.u., six bonds) and the bonds of twice the
nearest-neighbour length along $\langle111\rangle$ (9.44~a.u., eight bonds).
With $e_g=\{d_{z^2},d_{x^2-y^2}\}$ and $t_{2g}=\{d_{yz},d_{xz},d_{xy}\}$:
on the nearest-neighbour bond all elements are of similar size and the
$e_g$--$t_{2g}$ elements are finite; on the $\langle100\rangle$ bond only
diagonal elements survive and the $e_g$--$t_{2g}$ elements vanish, as
required by the fourfold axis along the bond (Fig.~\ref{fig:t2geg}); at
9.44~a.u.\ the $t_{2g}$ block dominates, with the opposite sign to the
nearest-neighbour shell.

\paragraph{FeSe (Fig.~\ref{fig:mat_fese}).} The site is Fe in the
nonmagnetic $\sqrt2a\times\sqrt2a\times c$ cell, and the orbitals refer to
the local axes of the Fe site (primed). The three strongest shells are the
Fe--Fe neighbours at $a/\sqrt2$, $a$ and $\sqrt2\,a$ (5.04, 7.12 and
10.07~a.u., four bonds each). On the nearest-neighbour bond, the susceptibility is large and can be of either sign (positive for $d_{y'z'}$ and negative for $d_{x'z'}$ and $d_{z'^2}$), so they partly cancel in the sum. On the second-nearest-neighbour bond,
the orbital contributions to the susceptibility are smaller, mostly negative and add up; this shell
therefore has the largest element sum and is the one quoted in the main text. On the first and third shells, the elements linking
$\{d_{z'^2},d_{x'^2-y'^2},d_{x'y'}\}$ with $\{d_{y'z'},d_{x'z'}\}$ vanish.

\section{Crystal structures}

\begin{table}[H]
\caption{Atomic positions of the structures in Table~I of the main text.}
\label{tab:positions}
\begin{ruledtabular}
\begin{tabular}{lll}
System & Space group & Wyckoff positions \\ \hline
bcc Fe, bcc Cr & $Im\bar3m$ & $2a$ $(0,0,0)$ \\
FeSe & $P4/nmm$ (origin 2) & Fe $2a$ $(\tfrac34,\tfrac14,0)$ \\
     &                     & Se $2c$ $(\tfrac14,\tfrac14,z)$, $z=0.2877$~\cite{Kumar2010} \\
2H-NbSe$_2$ & $P6_3/mmc$ & Nb $2b$ $(0,0,\tfrac14)$ \\
            &            & Se $4f$ $(\tfrac13,\tfrac23,z)$, $z=0.1183$~\cite{Johannes2006} \\
LaTe$_3$ & $Cmcm$ & La $4c$ $(0,y,\tfrac14)$, $y=0.1694$ \\
         &        & Te1 (LaTe layer) $4c$, $y=0.2953$ \\
         &        & Te2, Te3 (nets) $4c$, $y=0.5706$ and $0.9295$ \\
\end{tabular}
\end{ruledtabular}
\end{table}

\bibliographystyle{naturemag}
\bibliography{main}